\documentclass[letterpaper,twocolumn,10pt]{article}
\usepackage{usenix2019_v3}

\makeatletter
\let\@authorsaddresses\@empty
\makeatother

\AtBeginDocument{%
  \providecommand\BibTeX{{%
    \normalfont B\kern-0.5em{\scshape i\kern-0.25em b}\kern-0.8em\TeX}}}

\usepackage{cancel}
\usepackage{ulem}

\usepackage{xcolor}
\usepackage{bm}
\usepackage{enumitem}
\usepackage{amsmath,amsthm}
\usepackage[small,compact]{titlesec}

\DeclareMathOperator{\E}{\mathbb{E}}
\DeclareMathOperator*{\argmax}{arg\,max}

\usepackage{xspace}
\newcommand{\project}{\textit{Shockwave}\xspace}

\usepackage{cases}
\usepackage{subcaption}
\usepackage{caption}
\usepackage{ctable}
\usepackage{makecell}
\usepackage{tabularx}
\usepackage{newtxmath}
\usepackage{mathtools}
\usepackage{cleveref}
\usepackage{afterpage}

\usepackage{textcomp}

\crefformat{section}{\S#2#1#3}
\crefformat{subsection}{\S#2#1#3}
\crefformat{subsubsection}{\S#2#1#3}

\newtheoremstyle{mytheoremstyle}{5pt}{0pt}{\itshape}{}{\bf}{.}{.5em}{} 
\theoremstyle{mytheoremstyle}
\newtheorem{theorem}{Theorem}[section]
\newtheorem{corollary}{Corollary}[theorem]

\setlist[itemize]{leftmargin=3.5mm}

\allowdisplaybreaks
\usepackage{authblk}
\usepackage[T1]{fontenc}
\usepackage[utf8]{inputenc}
\makeatletter
\renewcommand\AB@affilsepx{, \protect\Affilfont}
\makeatother
\author[1]{Pengfei Zheng}
\author[1]{Rui Pan}
\author[2]{Tarannum Khan}
\author[1]{Shivaram Venkataraman}
\author[2]{Aditya Akella}
\affil[1]{University of Wisconsin-Madison}
\affil[2]{The University of Texas at Austin}

\begin{document}
\setlength{\abovedisplayskip}{0pt}
\setlength{\belowdisplayskip}{0pt}

\title{\project: Fair and Efficient Cluster Scheduling for Dynamic Adaptation in Machine Learning
}
\maketitle
\begin{abstract}
Dynamic adaptation has become an essential technique in accelerating distributed machine learning (ML) training. Recent studies have shown that dynamically adjusting model structure (e.g., lottery ticket hypothesis \cite{frankle2019stabilizing}) or hyperparameters (e.g., batch size \cite{agarwal2021adaptive}) can significantly accelerate training without sacrificing accuracy. However, existing ML cluster schedulers are not designed to handle dynamic adaptation. We show that existing schemes fail to provide fairness and degrade system efficiency when the training throughput changes over time under dynamic adaptation. We design \project, a scheduler with future planning that builds on two key ideas. First, \project extends classic market theory from static settings to dynamic settings to co-optimize efficiency and fairness. Second, \project utilizes stochastic dynamic programming to handle dynamic changes. We build a system for \project and validate its performance with both trace-driven simulation and cluster experiments. Results show that for traces of ML jobs with dynamic adaptation, \project improves makespan by 1.3$\times$ and fairness by 2$\times$ when compared with existing fair scheduling schemes.
\end{abstract}
\section{Introduction}
GPU-powered deep neural network (DNN) training is rapidly becoming a core workload in data centers~\cite{themis20,allox19, philly19}. Due to the sheer volume of training data and massive, ever-increasing model sizes, many DNN models cannot be trained on a single GPU device, and distributed, multi-GPU training has become the norm. The increasing demand for GPU devices motivates enterprises to consolidate their hardware resources and run their workloads in a shared GPU cluster~\cite{philly19}. Thus, building scheduling mechanisms that can fairly arbitrate among jobs competing for GPU resources and efficiently schedule them for high cluster utilization is important.  

While there has been a plethora of work in designing schedulers for DNN workloads, they do not use a rigorous approach to co-optimize system efficiency and fairness. Systems like Gandiva\cite{gandiva18} and Tiresias~\cite{gu2019tiresias} optimize makespan and average JCT (Job Completion Time) with techniques such as dynamic scaling, time-slicing, and over-subscription, but do not consider fairness.
Processor sharing~\cite{wierman2003classifying} based approaches such as DRF~\cite{drf11} and Gavel (Weighted Max-Min Fairness)~\cite{gavel20} provide instantaneous fair share of (dominant) resources in each scheduling round, but this can significantly undermine efficiency~\cite{altrustic16, optimus18}. 
Stride \cite{waldspurger1995lottery} scheduling-based approaches such as Gandiva-Fair~\cite{chaudhary2020balancing} require cluster operators to explicitly specify an individual job's share (e.g., A 20\% and B 80\% of GPUs), and manually specified fixed shares can violate long-term fairness for ML jobs~\cite{themis20}.
Finally, AlloX~\cite{allox19} and Themis~\cite{themis20} aim to provide long-term fairness by adopting a filter-based approach where within each round, a subset of jobs that are furthest from the fair share are filtered, and among the filtered jobs the ones which maximize efficiency are chosen by the scheduler. However, the filter value requires onerous hand-tuning; furthermore, even with careful tuning, using a fixed filter can lead to sub-optimal efficiency and fairness (\cref{sec:motivation}).

We design \project, a scheduler that leverages market theory to jointly optimize efficiency and fairness for ML training jobs in a systematic and principled fashion. We formulate a Fisher market~\cite{branzei2014fisher} where every job receives an equal budget to purchase resources from a central arbiter. The arbiter then computes prices such that the market reaches an equilibrium; i.e., each job's budget is spent to maximize its performance (e.g., training throughput) and all resources are completely sold. Formulating resource allocation using market theory is powerful because achieving market equilibrium guarantees both fairness and efficiency. Each job has equal purchasing power in acquiring resources, ensuring fairness.
Further, market-clearing equilibrium ensures work conservation and that each job's performance is maximized given its budget.

While  economic theory has been the basis of many prior systems (e.g., DRF~\cite{drf11}, Themis~\cite{themis20}, and REF~\cite{ref2014}), they all assume jobs have known static resource requests.
This assumption is no longer true for elastic ML training jobs~\cite{hwang2021elastic, qiao2020pollux, kungfu20} whose resource requirements dynamically change over time; further, the changes in resource requirements depend on model update patterns, and thus they are unknown apriori.  For example, training jobs can dynamically scale their batch size by computing the gradient
noise scale (GNS)~\cite{gns_scaling,kungfu20}. OpenAI has used batch size scaling (from 32 to 32M) to accelerate GPT-3 training by 500x~\cite{brown2020language} and similarly, BERT-Large training uses dynamic batch sizes (256 to 4096) to achieve a 2.5x speedup~\cite{qin2021simigrad}. 
In this paper, we extend market theory to develop an efficient and fair scheduler for ML jobs with elastic resource requirements.

Existing schedulers are either agnostic or reactive to dynamic changes and 
our experiments show (\cref{sec:motivation-uncertainty}) that
they fail to guarantee fairness or significantly degrade efficiency. 
The key reason for this is that an optimal schedule or weight assignment~\cite{chaudhary2020balancing} at the current instant can be suboptimal in the future, and reactively re-prioritizing jobs can be too late to compensate for the under-prioritization in the early phases.
State-of-the-art schedulers that accommodate dynamism, e.g.,  Pollux~\cite{qiao2020pollux} do so automatically on behalf of jobs, e.g., by automatically scaling batch sizes. We find that this can hurt training accuracy~\cite{agarwal2021adaptive, chin2019adascale} (\cref{subsec:support_user_customization_bs}); thus, our aim is to let users perform elastic changes as their algorithms demand. Achieving fair allocation under dynamism without assuming any control over said dynamism is challenging, and is not studied in existing research. We present a detailed comparison between \project and other schedulers such as Themis \cite{themis20}, AFS \cite{hwang2021elastic} and Pollux \cite{qiao2020pollux} in Section \ref{sec:motivation}.

To support dynamic changes in resource requirements over time, we extend the classic, static Fisher market and propose a new discrete-time, dynamic market that can ensure long-term efficiency and fairness. Using discrete time helps us capture the effects of running a market repeatedly over many rounds and a dynamic market helps us capture time-varying utility\footnote{A utility function maps a job's allocated resource (e.g., GPU) to the resulting performance (e.g., throughput).} for jobs. For example, consider a scenario where we are running 20 rounds of scheduling for a job. If a job's per-GPU batch size increases by $2\times$ after 10 rounds due to GNS scaling, its utility from being allocated one GPU ($u_0$) will also double after 10 rounds ($u_1 = 2u_0$). A static market will assume time-invariant utility, and will compute the accrued utility over 20 rounds for the job as $20u_0$; in contrast, a dynamic market can capture the change in utility for the job over time, and can accurately compute the accrued utility over 20 epochs as $30u_0$. Accurately computing the utility can enable the dynamic market to optimize fairness and efficiency over time. We prove that our dynamic market formulation (\cref{subsec:equilibrium_properties}) guarantees long-term efficiency and fairness properties such as maximized Nash social welfare over time, Pareto optimality over time, and sharing incentive. 

Implementing the dynamic market formulation in real systems is challenging for two main reasons.  First, the market formulation needs to know utility values in the future to compute market equilibrium. Dynamic adaptations in jobs are non-deterministically triggered, as they are dependent on gradient values that vary across models and datasets, which makes it challenging to predict utility in the future. Second, solving the dynamic market equilibrium for an (infinitely) long time horizon is difficult and impractical. It is computationally prohibitive and requires forecasting every job's future performance characteristics. Further, as jobs arrive and complete online, we need to periodically solve for the market equilibrium while maintaining low scheduling overheads.

To bridge the gap between theory and systems, \project addresses these challenges and implements a dynamic adaptation predictor and an approximate dynamic market. 
First, we observe that dynamic adaptation for real-world ML workloads follows a handful of patterns, and these patterns can be predicted using Bayesian statistics. We then develop methods to integrate these predictions into our dynamic market formulation.
Second, while performing round-based scheduling, we find that planning a schedule for an (infinitely) long time horizon can introduce significant overheads. To maintain low scheduling overheads, \project only plans the schedule for a finite length window (e.g, 30-60 minutes), and we design estimators that can capture the effects on long-term fairness and long-term efficiency that arise from short-term planning. This design helps us balance the system overheads without sacrificing long-term objectives.

We evaluate \project on a 32-GPU cluster testbed and use a simulator to study large-scale GPU clusters. Using multiple workloads derived from prior, real-world systems~\cite{gavel20, qiao2020pollux}, we find that \project improves makespan by 1.3$\times$ and fairness by 2$\times$ compared to existing fair DNN cluster schedulers including Themis\cite{themis20}, Gavel \cite{gavel20}, AlloX \cite{allox19}, etc. We further evaluate \project on differently sized clusters. Using a simulator built with the same scheduler as in a physical cluster we find that \project scales to schedule 900 active DNN training jobs on 256 GPUs and maintains the benefits in makespan (1.26-1.37$\times$) and fairness (2.5-3.1$\times$) when compared to existing schedulers. We show that our solver overhead remains low and is less than 12.5\% of a two-minute-long round duration.\footnote{The solver runs asynchronously in a separate thread and does not block the main scheduling loop.} \project is open sourced at \url{https://github.com/uw-mad-dash/shockwave}.

\section{Motivation}
\label{sec:motivation}
We begin by motivating the need to design a new cluster scheduler for machine learning workloads.

    \begin{table}[h!]
    \small
    \centering
    \begin{tabularx}{\columnwidth}{c c c c c}
        \hline
        \makecell[c]{Filter $f$} & \makecell[c]{Worst FTF-$\rho$} & \makecell[c]{SI} & \makecell[c]{Avg. JCT} \makecell[c]{Makespan} \\ 
        \hline
        \makecell[c]{Adaptive - $1/\frac{1}{3}$/$\frac{2}{3}$} & \makecell[c]{0.83} & \makecell[c]{\checkmark} & \makecell[c]{5} \makecell[c]{7} \\\hline
        \makecell[c]{Fixed - 1/3} & \makecell[c]{1.0} & \makecell[c]{\checkmark} & \makecell[c]{5.7} \makecell[c]{7} \\\hline
        \makecell[c]{Fixed - 2/3} & \makecell[c]{1.1} & \makecell[c]{$\times$} & \makecell[c]{5.7} \makecell[c]{7} \\\hline
        \makecell[c]{Fixed - 1} & \makecell[c]{1.1} & \makecell[c]{$\times$} & \makecell[c]{6.0} \makecell[c]{7} \\ \hline
    \end{tabularx}
    \caption{\label{tab:example_filter_1} Themis example: using a fixed filter yields suboptimal JCT and/or fairness compared with an adaptive filter. Figure~\ref{fig:themis-1/3-example} visualizes the schedule for $f=2/3$, showing the cluster and job setting, and demonstrates how a filter works.
    }
\end{table}\vspace{-1.5em}
 \begin{figure}[!h]
    \centering
    \includegraphics[width=0.48\textwidth, trim=0.1cm 0.5cm 0.1cm 0.1cm,clip]{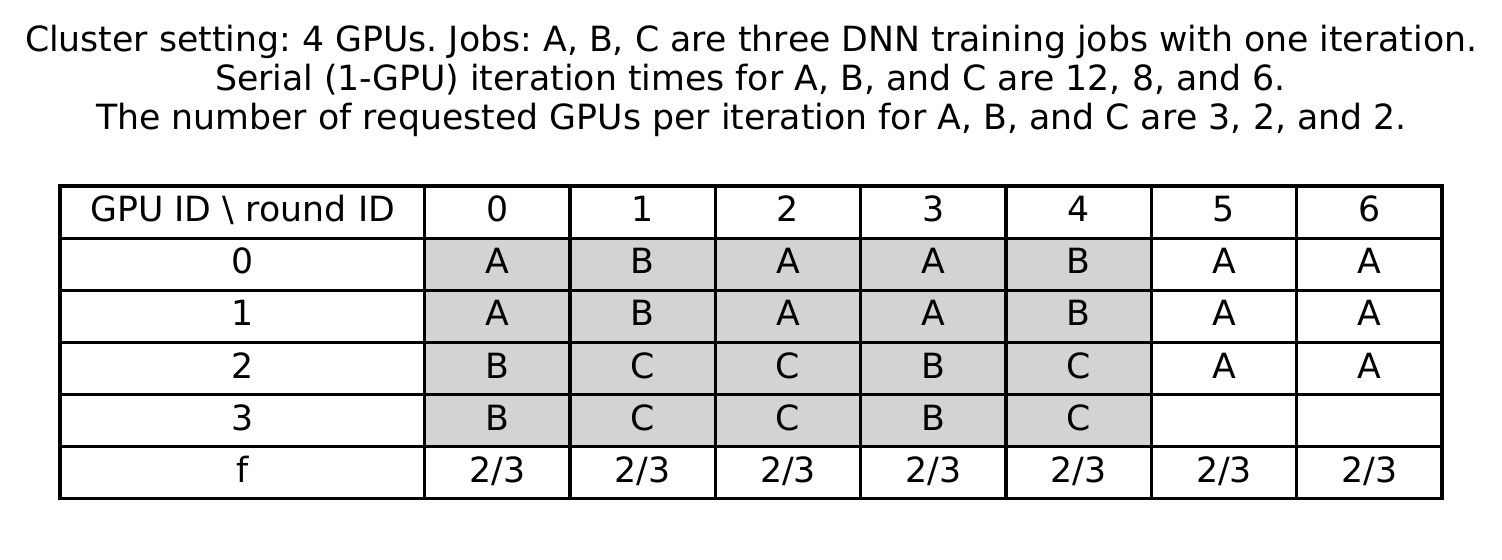} 
    \caption{Example - Themis \cite{themis20} with a static filter ($f=2/3$). In each round of allocation, the filter (grey color) selects $2/3$ of the jobs unfairly treated so far. The resulting FTF-$\rho$ values for jobs (A, B, C) are (0.78, 0.83, 1.1), showing a static filter hurts fairness. As in Themis, we assume a linear slowdown when the number of allocated GPUs is less than requested.}
    \label{fig:themis-1/3-example}
    \vspace{-0.1in}
\end{figure}
\subsection{Jointly Optimizing Fairness and Efficiency}
\label{subsec:joint_motivation}
Existing scheduling mechanisms lack a systematic approach to jointly optimize fairness and efficiency. We first formally define a \emph{fairness} metric: we adopt the definition of fairness used in recent work such as Themis~\cite{themis20}, Gavel~\cite{gavel20}, and Pollux~\cite{qiao2020pollux}: Finish Time Fairness (FTF) $\rho(G)=\frac{t_{schedule}}{t_{egalitarian}}$; where $t_{schedule}$ represents the job finish time resulting from a policy $G$, and $t_{egalitarian}$ is $t_{exclusive}\cdot N$; $N$ indicates the number of contending jobs, $t_{exclusive}$ indicates run time when running exclusively with requested resources. FTF $\rho>1$ ($\rho<=1$) indicates that a job is unfairly (fairly) scheduled.  
Note that while we focus on FTF, \project{}'s underlying market formulation can be extended to support other fairness metrics. For example, by assigning different budgets to jobs we can support weighted proportional fairness \cite{kushner2004convergence} with the budgets encoding priorities.

Second, we define a policy as \emph{efficient} if it minimizes makespan, or correspondingly, maximizes cluster utilization given a sequence of jobs.

\noindent \textbf{Instantaneous fair-share sacrifices efficiency.} Existing fair sharing policies, such as Processor-Sharing (PS)~\cite{wierman2003classifying} and its multi-dimensional extension, Dominant Resource Fairness (DRF)~\cite{drf11}, guarantee that at each instant, any job in the schedule obtains exactly a $1/N$ share of the (dominant) resources. However, restricting schedulers to instantaneous fair share can adversely degrade long-term efficiency. Previous work in Altruistic Scheduling (AS)~\cite{altrustic16} has shown that sacrificing instantaneous fair share and letting some jobs altruistically contribute resources can improve efficiency by 26\%~\cite{altrustic16}.

\noindent \textbf{Using filters to balance fairness and efficiency is sub-optimal.}
Given the limitations of instantaneous fair sharing schemes, recent work~\cite{allox19, themis20} has proposed using round-based schemes that optimize instantaneous efficiency and long-term fairness. 
Within each round of scheduling, AlloX~\cite{allox19} and Themis~\cite{themis20} select for allocation a fixed fraction ($f$) of jobs that have attained the least resource share in the past. Within these filtered jobs, the scheduler tries to maximize efficiency. Across rounds, the filter compensates for jobs unfairly scheduled in the past and thus pursues fairness in the long run. 

Existing studies pre-specify a fixed value for filter $f$ across rounds, but we find that adopting a fixed filter can incur a loss in average JCT or makespan~\cite{themis20}, and filter tuning is challenging. Table~\ref{tab:example_filter_1} uses a simple example with three jobs to show how different filters yield very different performance outcomes:  fixed filter values $f=1$ and $f=\frac{2}{3}$ violate finish time fairness ($\rho > 1$) while $f=1/3$ leads to worse JCT. We included the full toy examples in Appendix~\ref{apdx:toy_example}.
Tuning the hand-crafted fairness filter is challenging without any insight into the resulting performance outcomes, and it is more difficult when the workload varies. %

Overall, this argues for a rigorous, systematic approach that jointly and intrinsically (i.e., without knob-tuning) optimizes efficiency and fairness in resource sharing. %

\begin{figure*}[htbp!]
\centering
\begin{subfigure}{.3\textwidth}
  \centering
  \includegraphics[width=\textwidth, trim=0.2cm 0.2cm 0.2cm 0.2cm,clip]{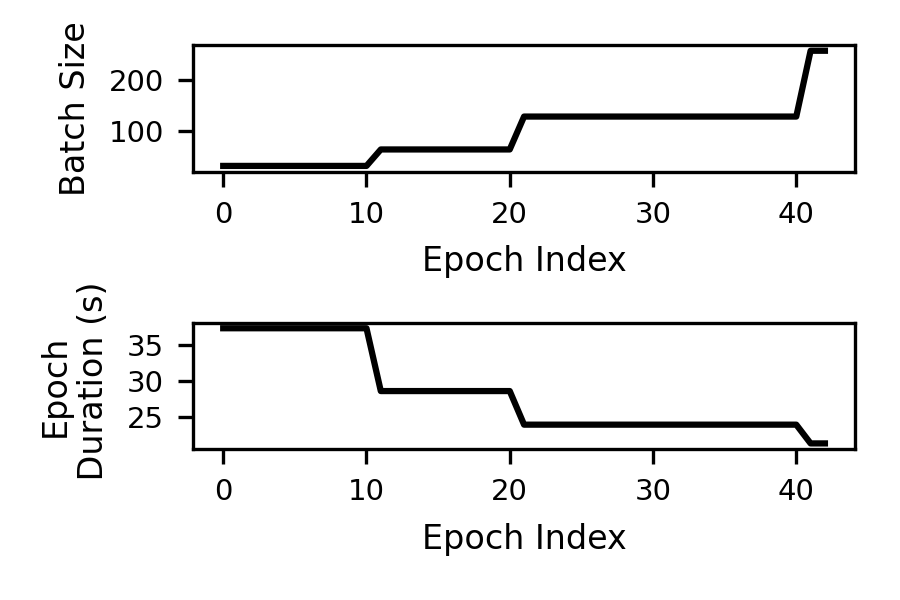}\caption{Dynamic Adaptation}
  \label{fig:mv-example-adaptation}
\end{subfigure}
\begin{subfigure}{.3\textwidth}
  \centering
  \includegraphics[width=\textwidth, trim=0.1cm 0.0cm 0.1cm 0.1cm,clip]{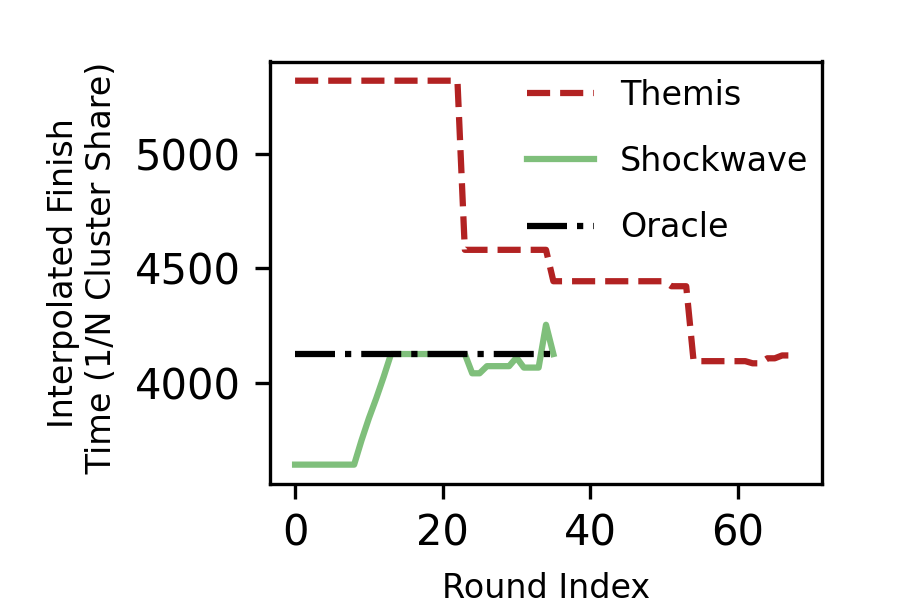}  
  \caption{$T_{egalitarian}$ - Interpolated finish time under $1/N$ cluster share}
  \label{fig:mv-example-interpolation}
\end{subfigure}
\begin{subfigure}{.3\textwidth}
  \centering
  \includegraphics[width=\textwidth, trim=0.1cm 0.0cm 0.1cm 0.1cm,clip]{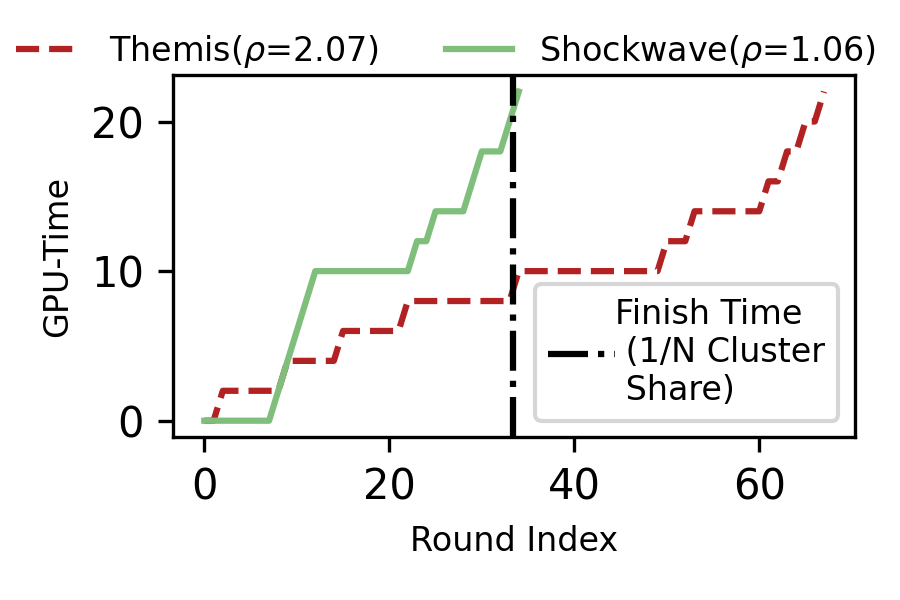}  
  \caption{GPU Allocation}
  \label{fig:mv-example-visshed}
\end{subfigure}
\caption{Example - Reactive scheduling (Themis \cite{themis20}) for dynamic adaptation breaks finish time fairness. Proactive scheduling (\project) for dynamic adaptation preserves finish time fairness.}
\label{fig:motivation-ftf-adapation}
\vspace{-0.1in}
\end{figure*}

\subsection{Handling Dynamic Batch Size Scaling}
The above goal of optimizing for fairness and efficiency is made further challenging in the presence of dynamism.
Dynamism can result in a number of different scenarios. For example, dynamism can result from job arrivals leading to time-variant cluster contention, and systems like AFS~\cite{hwang2021elastic} are designed to improve JCT by adjusting shares based on job arrivals.
On the other hand, dynamism can arise from training jobs that can change their training configurations dynamically. For example, if a job uses gradient noise scale (GNS)~\cite{gns_scaling,kungfu20}, the batch size used can change dynamically.
This can affect fair scheduling because when a job switches to using a large batch size, the per epoch time will decrease, and thereby its remaining running time will also decrease (Figure~\ref{fig:motivation-ftf-adapation}(a)). 
 Unlike prior systems which only handle dynamism that arise from job arrivals, \project (and prior work in Pollux~\cite{qiao2020pollux}) focus on handling dynamic changes in batch sizes of training jobs. %

\noindent \textbf{Being agnostic or reactive to dynamic adaptation breaks finish time fairness.} We show that being agnostic or reactive to dynamic changes (or dynamic adaption) can yield severe unfairness. Finish time fairness (FTF) implies a soft deadline $t_{egalitarian}$ = $t_{exclusive}\cdot N$ for job completion; the later the job finishes after the deadline, the more unfair the schedule is. 
Computing FTF requires computing the exclusive run time (i.e.,$t_{exclusive}$), which is straightforward for static jobs since run time roughly equals training throughput (samples per second) times the remaining amount of work (remaining number of epochs).
However, for jobs that use dynamic adaptation, future training epochs could be significantly faster because of using a larger batch size. Agnostic scheduling and reactive scheduling are both unaware of future speedups and hence overestimate run time, and thus mistakenly extend the deadline $t_{egalitarian}$ leading to significant unfairness.

Figure \ref{fig:mv-example-interpolation} uses a job from our trace to show the difference between Themis, which uses a reactive approach, and \project, which uses a proactive approach. The job dynamically doubles batch size three times from 32 to 256, and gradually boosts training speed by up to 1.7$\times$ (Figure \ref{fig:mv-example-adaptation}). Themis is notified and updates the job's throughput immediately after each batch size scaling, and recomputes the estimated finish time based on that (red dashed line in Figure \ref{fig:mv-example-interpolation}). Changes in the estimated finish time lead Themis to detect that the job has received less than its fair share and Themis attempts to prioritize this job in future scheduling rounds. However, the job has already suffered from under-prioritization in its early phases and misses the fairness deadline by 2.07$\times$ (Figure \ref{fig:mv-example-visshed}). 
Agnostic scheduling is even worse and increases the job's FTF $\rho$ to 3.07; we omit this from the figure.

\noindent \textbf{Being agnostic or reactive to dynamic adaptation degrades system efficiency.} 
Many makespan-minimizing algorithms, such as Mixed Integer Linear Programming (MILP), Longest Processing Time (LPT)~\cite{della2020longest}, and JCT (Job Completion Time) minimization algorithms such as Shortest Remaining Time (SRPT) and AlloX \cite{allox19}, rely on the exact job run time, or the ordering of jobs' run time to derive a schedule.

\begin{figure}[!t]
  \centering
  \includegraphics[width=0.85\linewidth]{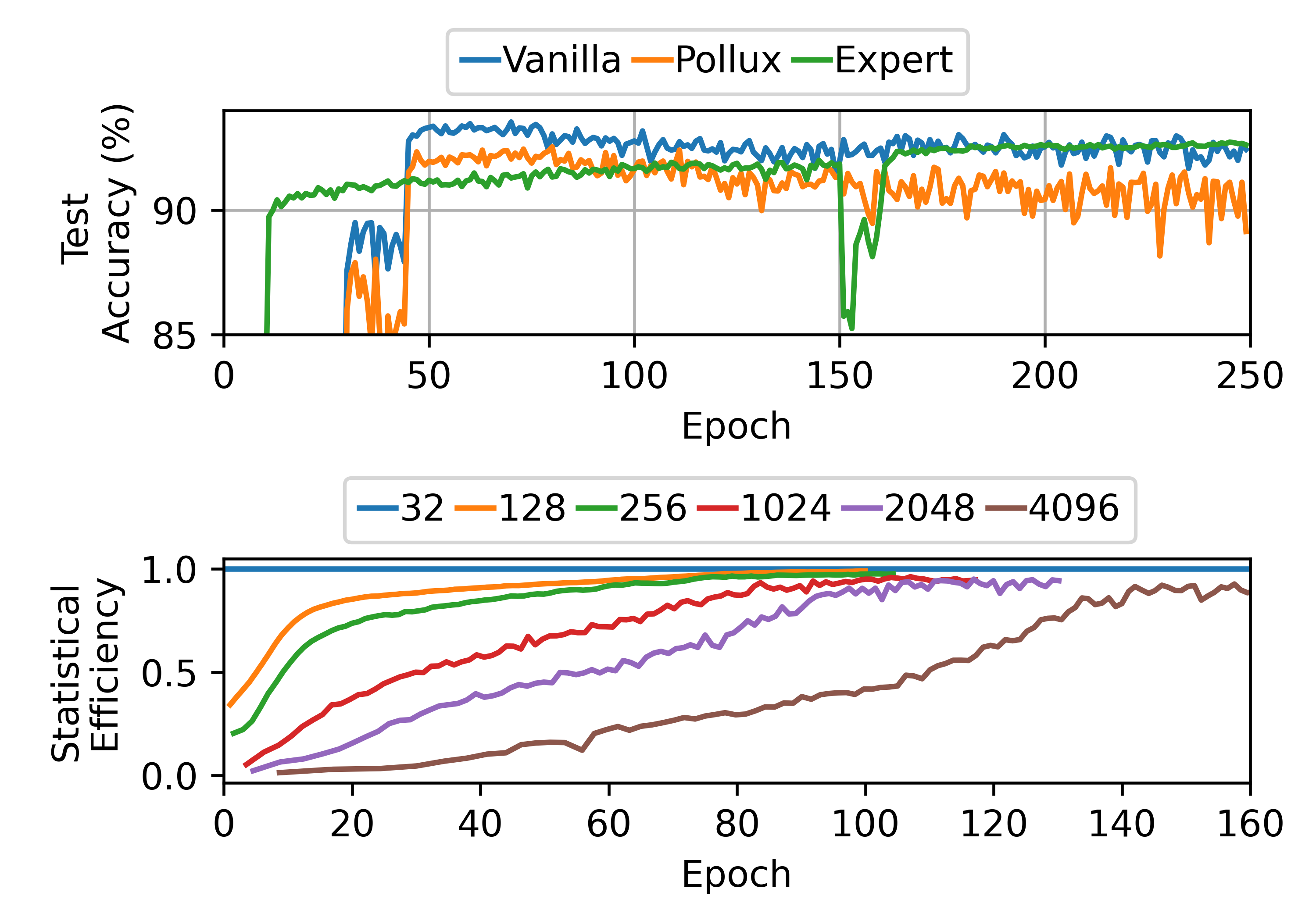}
  \caption{Comparing model accuracy (ResNet18-CIFAR-10) for vanilla training, expert-set batch size scaling, and Pollux autoscaling. The legends in the bottom figure indicate batch size.} 
  \label{fig:mv-example-accuracy_gap_cifar10}
  \vspace{-0.1in}
\end{figure}

\begin{figure}[!t]
  \centering
  \includegraphics[width=\linewidth]{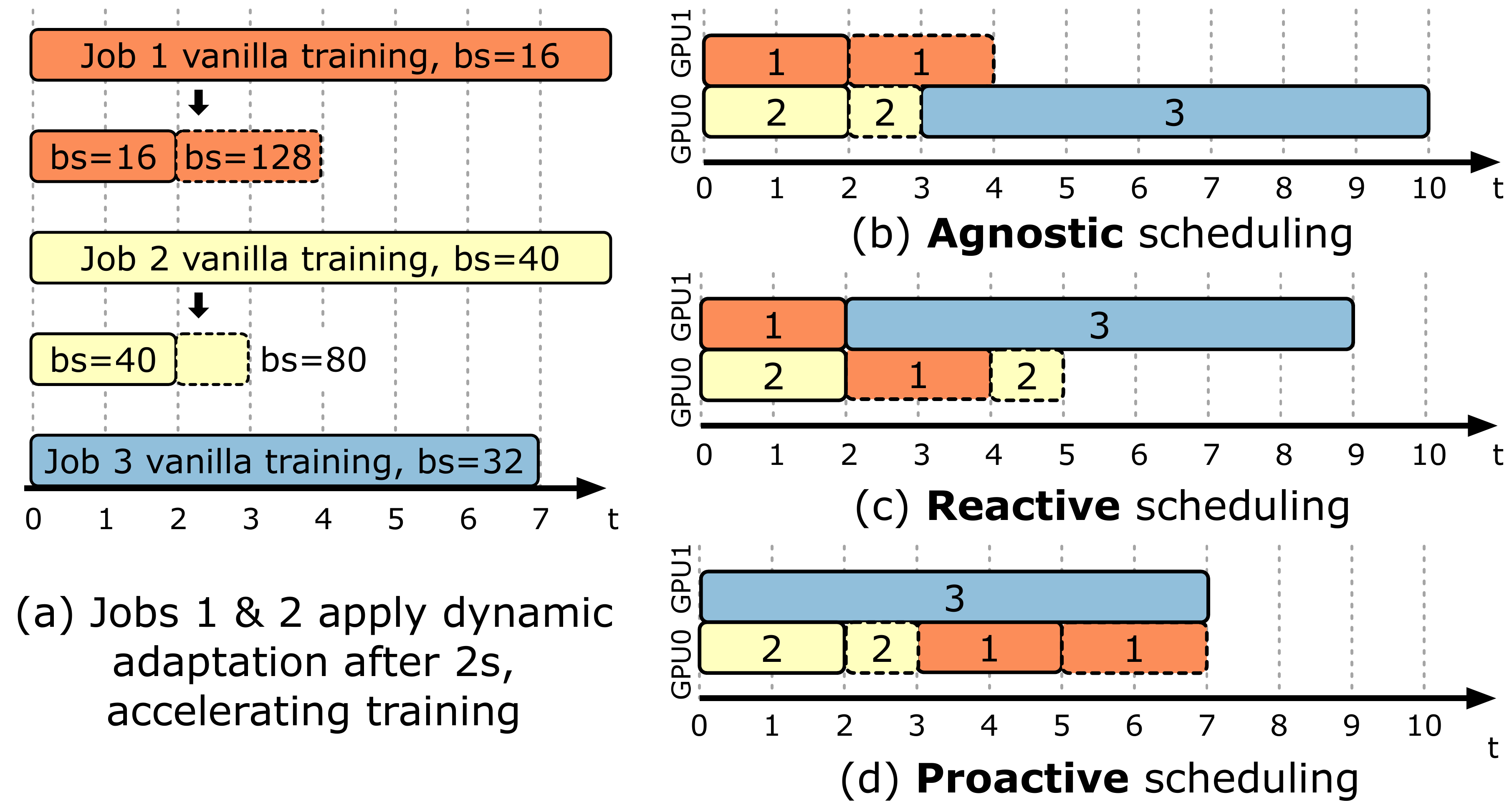}
  \caption{Being agnostic and/or reactive to dynamic adaptation undermines efficiency while proactive scheduling minimizes makespan and maximizes efficiency.} 
  \label{fig:mv-example-efficiency-adaptation}
  \vspace{-0.1in}
\end{figure}

However, dynamic adaption adaptively changes a job's throughput, and thus a job's run time can be reduced (when batch size is increased) or prolonged (when batch size is decreased) on the fly. This means that when making scheduling decisions, the job run time estimated using initial or current throughput is only valid for the current instant, and if it is used beyond that system efficiency can be significantly undermined. In Figure~\ref{fig:mv-example-efficiency-adaptation}, the example shows that for MILP makespan minimization, being reactive to dynamic adaptation yields a 22.3\% worse makespan and 28\% worse cluster utilization compared to proactive scheduling. Reactive scheduling considers $J1$ and $J2$ as long-running jobs from their initial throughput and prioritizes them to minimize makespan. But due to dynamic adaptation, $J1$ and $J2$ become shorter than $J3$ in their second epoch, and it is too late to compensate and re-prioritize $J3$. Being completely agnostic to dynamic adaption is even worse, yielding a 30\% worse makespan.

Overall, the above points motivate the need for a scheduler that can model future dynamic adaptation and account for this uncertainty while optimizing for both fairness and efficiency.

\subsection{Supporting User-defined Dynamic Ddaptation}\label{subsec:support_user_customization_bs}

While dynamic adaptation with batch size scaling is a key enabler for efficient training of large-scale DNNs, improper changes to the batch size can adversely impact convergence properties and degrade model accuracy. Thus, unlike systems such as Pollux~\cite{qiao2020pollux} which automatically modify the batch size of training jobs, we argue that schedulers should support user-defined dynamic adaptation schedules to avoid affecting training accuracy. This is mainly because no adaptive batch size scaling technique works consistently well across all datasets and optimizers. As a result, ML researchers have developed many different batch sizing scaling policies including linear scaling rule~\cite{brown2020language}, Accordion~\cite{agarwal2021adaptive}, Gradient Noise scale (GNS)~\cite{qiao2020pollux}, SimiGrad \cite{qin2021simigrad}, Hessian eigenspectrum~\cite{yao2018hessian}, etc.

We next study an example of how choosing an incorrect batch size schedule can affect accuracy.
In Figure~\ref{fig:mv-example-accuracy_gap_cifar10}, we consider a CIFAR-10 training job on 2 GPUs with ResNet18 with an initial batch size of 32. When using Pollux~\cite{qiao2020pollux}, the end-to-end training time reduces by $5\times$ as Pollux scales up the batch size from 32 to 64 at epoch 1, and then up to 314 at epoch 2, then to 690 at epoch 30, and finally up to 1682 at epoch 70 till completion. However, this aggressive scaling leads to a 2-3\% accuracy loss. Plotting the statistical efficiency (as defined in Pollux~\cite{qiao2020pollux}), we find that using large batch sizes in the first 30 epochs leads to accuracy degradation. 
Our conversation with the authors of Pollux suggests that the accuracy degradation depends on the initial batch size used (32 in this case) and can thus vary across jobs.\footnote{We also found that the statistical efficiency metric in Pollux can be incorrect for Neural-MF models~\cite{ncf17}. We include details of this experiment in Appendix~\ref{apdx:accuracy_gap}.}

We also tested an expert heuristic for batch size scaling of ResNet18 training on CIFAR-10. The heuristic scales up the batch size when the gradient norm\cite{agarwal2021adaptive} has insignificant (<50\%) changes, and does not scale in the initial epochs 20 epochs and the 10 epochs before and after each learning rate decay. This expert-defined scaling schedule has minimal accuracy loss and is $3\times$ faster than vanilla training. Such expert heuristic and the associated thresholds vary across models and datasets; the above heuristic is specific to ResNet-18-CIFAR-10 training and is not easily transferable. For example, for ResNet-50-ImageNet training, the experts propose a different heuristic that scales up the batch size by a factor of ten at the 30th, 60th, and 80th epoch, respectively ~\cite{increasebs18}. Thus, while prior work has developed scheduling mechanisms for specific 
dynamic adaptation techniques, in \project, on the other hand, we assume no preference for any technique and respect any choice made by users regarding how to dynamically scale a training job’s batch size.

In summary, we find that automatically performing dynamic adaptation runs the risk of accuracy degradation. Hence in this work, we aim to develop a scheduler that can observe and forecast future scaling events but treats dynamic adaptation as a part of the user's program that cannot be modified.

\section{Overview}
We next present an overview of \project{}, a new scheduling framework that jointly optimizes efficiency and fairness for machine learning workloads in shared clusters.

\noindent\textbf{Using Market Theory for Efficiency and Fairness} 
In \project{} we propose using market theory to provably guarantee efficiency and fairness for resource sharing. While prior schedulers~\cite{themis20, zahedi2018amdahl} have also leveraged market theory for fair sharing, they are built on static market models which assume that resource requests for a job don't change over time. We find that the fairness and efficiency guarantees of a static market do not hold when jobs dynamically change over time~\cite{fikioris2021incentives}. 
Thus, \project extends the classic, static market to a discrete-time, dynamic market, to support efficient and fair resource sharing under dynamic adaptation.

\noindent\textbf{Predicting Dynamic Adaptation}\label{sec:motivation-uncertainty}
Building a dynamic market alone is not enough as it presumes perfect future knowledge of jobs' dynamic adaptation behavior; that is, the market needs to know when and how much jobs' performance (e.g., training throughput) is changed by dynamic adaptation as training progresses. As ML training itself is a stochastic process, the trajectory of dynamic scaling is intrinsically uncertain. We address this problem in \project by forecasting the trajectory of dynamic adaptation and developing methods to use these forecasts in the dynamic market.

\noindent\textbf{Scalable System Implementation}\label{sec:contribution_approxi}
Solving a dynamic market and predicting dynamic adaptation introduces scheduling overhead. We build a centralized, round-based scheduler~\cite{gavel20} and incorporate tractable approximations that can ensure the overhead remains low even as we scale the number of GPUs and cluster size. We find that \project{} can maintain low overhead while scheduling every 120 seconds and scale to handle 900 active jobs running on a cluster of 256 GPUs.

\section{Dynamic Market Theory Formulation}
We begin by describing our theoretical formulation of a discrete-time, dynamic market and the properties it provides.

\subsection{Volatile Fisher Market}
Market theory provides a fundamental approach to provably guarantee efficiency and fairness in resource sharing. The equilibrium of a \emph{Fisher Market}~\cite{branzei2014fisher}, which is solved by maximizing Nash Social Welfare (NSW)~\cite{caragiannis2019unreasonable}, is a strong condition that implies all fairness guarantees used in prior systems. It is known that Fisher market equilibrium (under equal endowment) implies Pareto Optimality (PO), Envy-freeness (EF), and Proportionality (PR), which are fairness properties adopted by existing systems like DRF~\cite{drf11}. 

We define efficiency in terms of the utility of a job, where \emph{utility} is a function that maps allocated resources to the resulting job progress (e.g., throughput improvement if we allocate more resources). The market equilibrium for a Fisher market has also been shown to maximize efficiency~\cite{nsw17}. Thus, we explore the applicability of Fisher markets for DL jobs.

\noindent\textbf{From static market to dynamic markets: Volatile Fisher Market.} Classic Fisher Market assumes static, time-invariant utility for jobs, and a recent study \cite{fikioris2021incentives} shows that efficiency and fairness guarantees can be violated for dynamic, time-variant utilities.
Prior work \cite{azar2010allocate, angelopoulos2005line} on dynamic markets has also studied settings where goods (resources) arrive online, while our market considers  a different setting where buyers in the market have time-variant utilities over goods.

To perform efficient fair sharing under dynamic adaptation, we extend the static Fisher market to a discrete-time, dynamic market. We name this new market \textbf{Volatile Fisher Market (VFM)}. We prove that maximizing Nash Social Welfare Over Time (i.e., $\mathsf{NSW_{OT}}$ in Equation \ref{eq:nswot}) solves the market equilibrium of VFM and establishes long-term efficiency and fairness properties, such as Proportionality Over Time, i.e., $\mathsf{PR_{OT}}$, which has strong implications for finish time fairness and sharing incentive. We leave the formulation and related proofs of VFM in Appendix \ref{apdx:vfm_formulation}-\ref{apdx:theorem_eg2equilibrium}, and provide a succinct description below.

VFM operates at discrete time intervals $t=1,\dots, T$. At each time instant, a central seller (the scheduler) sells resources (e.g., GPUs and/or CPUs) to buyers (jobs). All resources are volatile. That is, resources bought by a job at time $t'$ cannot be carried over to the future time steps $t>t'$. To model dynamic adaptation, the utility for any job $i$ is a sequence of time-variant functions $u_{it}$ ($t=1, \dots, T$). For example, a job might have a utility $u_{0}$ when the batch size is $16$ and its utility could double $u_{1}=2u_{0}$ when the batch size doubles at $t=1$. Since jobs' utilities can change over time, this creates dynamic changes in demands over time, and thus, resource price, which is achieved at equilibrium, is also time-variant. We assume that each job is endowed with an initial budget to spend across rounds. The budget for a job reflects its purchasing power and different budgets can reflect scheduling priority.

Given the resource demands, budget, and utility for each job, at every time instant, the VFM solves for an allocation and assignment of prices that can lead to \emph{market equilibrium}. We define the market to have reached an equilibrium when two conditions are satisfied. \textbf{(a) Optimal Spending}: Each job's utility accrued over time, i.e, $\sum_{t=1}^{T} u_{it}$, is maximized under its budget. \textbf{(b) Work-conserving}: There are no leftover resources if the price for the resources is non-zero.

\subsection{Equilibrium Properties}\label{subsec:equilibrium_properties}
The market equilibrium achieved by VFM has a number of powerful properties that we define below. Proofs for them are in Appendix \ref{apdx:vfm_formulation}-\ref{apdx:theorem_ftf}.

\noindent\textbf{Cluster-level performance.} The equilibrium of VFM maximizes Nash Social Welfare Over Time($\mathsf{NSW_{OT}}$) which is an indicator of cluster-level performance. 
\begin{equation}\label{eq:nswot}
\begin{gathered}
    \mathsf{NSW_{OT}}(U_1(\bm{X_1}), \dots, U_N(\bm{X_N}))=\prod_i U_i(\bm{X_i})^{\frac{B_i}{\sum_iB_i}},\;\;\\
    U_i(\bm{X_i})=\sum_t \bm{u_{it}}(x_{it})
\end{gathered}
\end{equation}

Let $U_i(X_i)=\sum_t u_{it}(\bm{x_{it}})$ represent the utility (e.g., epoch progress) for a job $i$ accrued over rounds $t=1,\dots, T$, $\bm{X_i}$ represent the sequence of allocations $\bm{x_{i1}}$, $\dots$, $\bm{x_{it}}$, $\dots$, $\bm{x_{iT}}$ received for individual rounds, and $B_{i}$ represent the budget provided for the job. Maximized $\mathsf{NSW_{OT}}$ guarantees that the (weighted) geometric mean of job progress is maximized for a $T$-round-long time horizon. In effect, this property guarantees that the overall cluster-wide utility is maximized across all jobs, thus leading to improved utilization.  %

\noindent  \textbf{Pareto Optimality over time.} We prove that maximized $\mathsf{NSW_{OT}}$ also implies Pareto Optimality Over Time ($\mathsf{PO_{OT}}$), which guarantees resource allocation efficiency. Specifically, $\mathsf{PO_{OT}}$ ensures that each job has no surplus resources at each instant; i.e., we cannot increase one job's training progress without depriving that of another job.

\noindent \textbf{Finish Time Fairness (FTF) over time.} We also show that maximized $\mathsf{NSW_{OT}}$ minimizes the product of FTF across all jobs and that this directly leads to sharing incentive (assuming budgets are equal). A formal statement is in Corollary~\ref{theorem_ftf} and the proof is in Appendix~\ref{apdx:theorem_ftf}.

\begin{corollary}\label{theorem_ftf}
The equilibrium of the Volatile Fisher Market with linear or Leontief utility at each instant (a) minimizes the product of FTF ($\rho$) across all jobs (i.e., $\prod_i \rho_i$); (b) when the budgets assigned to jobs are equal, the equilibrium provably guarantees Sharing Incentive (SI), i.e., all jobs' FTF $\rho$ are no greater than 1, i.e., $\rho_i\leq 1$,\, $\forall i$.
\end{corollary}

Thus, our formulation of volatile Fisher markets can capture time-varying utility for jobs while providing a number of powerful guarantees in terms of fairness and efficiency.

\subsection{Handling Uncertainty}
The Volatile Fisher Market model described above assumes perfect knowledge of the future. That is, the model requires knowing at which time point $t$ will the throughput change due to dynamic adaptation. However, dynamic adaptation in jobs is non-deterministically triggered, as they are dependent on stochastic gradient values and can thus vary across models and datasets. %
To handle this, in \cref{sec:regime_model}, we develop methods to predict dynamic adaptation in jobs. But given that the predictions are random variables, we further extend our above formulation to derive a VFM that can handle uncertainty in future request demands. We show that this extension guarantees Maximized Nash Social Welfare Over Time in Expectation ($\mathsf{MNSW_{OTE}}$). Details are provided in Appendix~\ref{apdx:egprogram_stochastic}.

\section{Predicting Dynamic Adaptation}\label{sec:regime_model}

In this section, we develop techniques to predict the dynamic adaptation of the batch size that occurs in elastic ML training workloads. Our key insight in developing a predictor is to leverage our knowledge about techniques that are used for batch size scaling~\cite{gns_scaling, agarwal2021adaptive} and thereby restrict the search space of possible batch size changes. We next define how changes in batch size over time can be viewed as trajectories and then describe how we can use Bayesian statistics to predict which trajectories are most likely.

\noindent\textbf{Dynamic adaption, regimes, and randomness.} We define a regime of training $R$ as a tuple $R=(c,f)$; where $c$ indicates the job configuration (e.g., batch size) used in the regime and $f$ represents the duration (as a fraction of the total epochs) that this regime lasts. For example, if a 100-epoch-long DNN training job starts with batch size 32 (denoted $BS_{32}$) for epoch 1-20, then the first regime is denoted as $c1=BS_{32},f1=0.2$. We define a \emph{trajectory} as a sequence of regimes. For example, if the same job scales up to $BS_{64}$ for epoch 21-80, and finally scales down to $BS_{32}$ for epoch 81-100, then its trajectory is represented as $(c1=BS_{32}, f1=0.2)$$\rightarrow$$(c2=BS_{64}, f2=0.6)$$\rightarrow$$(c3=BS_{32}, f3=0.2)$. Thus, given a new job, each regime $c_i, f_i$ is a tuple of \underline{random} variables.

\noindent\textbf{Leveraging domain knowledge.} We leverage domain knowledge about techniques used for batch size scaling to constrain the random variables. Techniques for scaling batch size have deterministic patterns. (a) \textbf{Accordion~\cite{agarwal2021adaptive}} only alternates between two configurations $c_1$ for small batch size and $c_2$ for large batch size.  When gradient values change slowly (below a threshold) during training, Accordion scales up the batch size from $c_1$ to $c_2$, and when gradient values change rapidly, Accordion scales from $c_2$ back to $c_1$. (b) \textbf{GNS~\cite{gns_scaling}} only scales up the batch size up to the pre-specified limit and never scales down. Existing studies show that the gradient noises tend to grow throughout training, implying that GNS will gradually scale up the batch size and never scale down~\cite{gns_scaling,kungfu20}. We use a simple model of GNS scaling where, as the gradient noise grows above a relative threshold, the batch size doubles.

We choose Accordion and GNS as representative batch size scaling patterns as they have been used in prior systems like KungFu~\cite{kungfu20} and Pollux~\cite{qiao2020pollux}.
Further, their scaling decisions are completely determined by gradient states (i.e., gradient norms and noises), which encode the stochasticity induced in back-propagation algorithms. Most other dynamic batch sizing policies also adapt to gradient states and we plan to add support for more policies in the future.

\noindent\textbf{Prior for regime transition.} Given that batch size scaling rules have deterministic configuration transitions, the only random variable is a regime's duration. For a job with $K$ regimes, we define a probabilistic model $P(f_1,\dots,f_K)$ to represent the probability that regime $k$ ($k$=$1,\dots,K$) lasts for $f_k$ fraction of epochs. We also note that the sum of all regimes' epoch fractions needs to sum up to 1 \footnote{We don’t make any stationary assumptions about the distribution.}.
Given this formulation, we use an approach based on Bayesian statistics to predict regime duration. At a high level, our approach is to define a prior distribution of regime duration and then update the posterior distribution in real time as training progresses. The key challenge here is in determining how we can update the posterior as training progresses.

\begin{figure}[!tb]
  \centering
  \vspace{-0.1in}
  \includegraphics[width=0.75\linewidth]{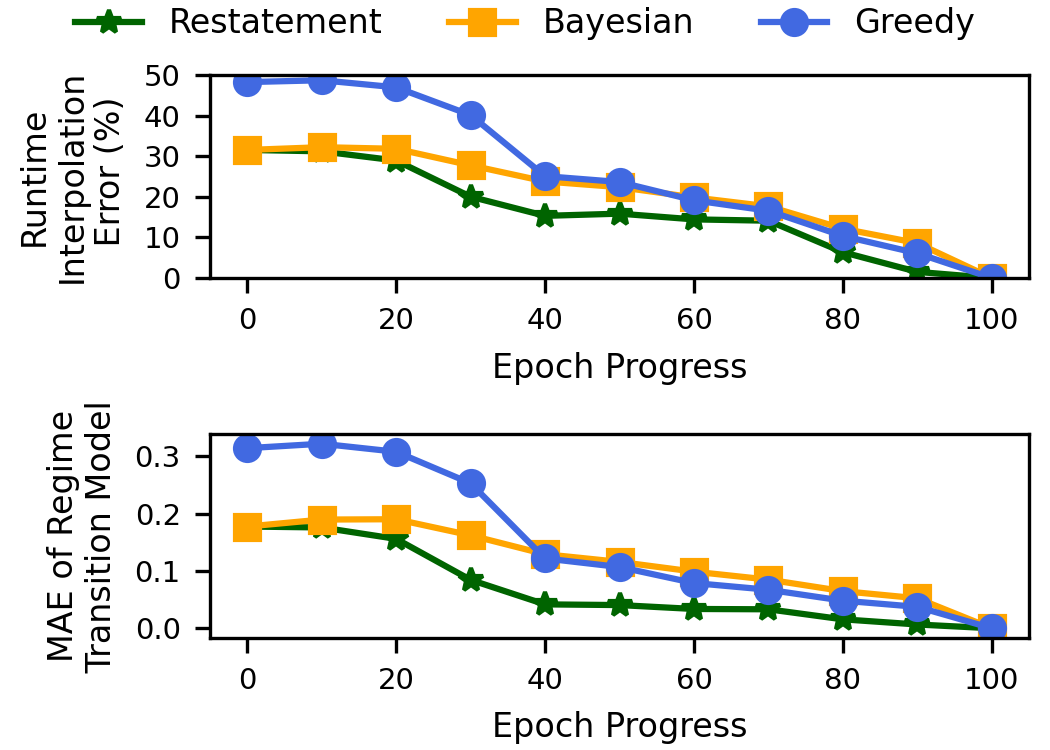}
  \caption{\project Dynamic Adaptation Modeling Error.} \label{fig:dirichlet}
\end{figure}

\noindent \textbf{The restatement posterior update rule.} 
Given our problem formulation, we adopt the commonly used Dirichlet prior $Dir(n1,\dots,\allowbreak n_K)$. %
A standard Bayesian posterior update rule assumes the epoch samples of individual regimes are independently and randomly drawn as training progresses. But this does not hold in practice. Epochs of the $k$-th regime can only emerge if the $k-1$-th regime finishes. To deal with the temporal-dependence issue, we design a simple update rule, named the \emph{restatement rule}, for posterior updates. The restatement rule only updates the prior's parameters that correspond to completed epochs, while continuing to believe that the ongoing and future regimes will evenly split the remaining epochs. Specifically, suppose a user specifies that at a maximum, $K$ regimes can exist, the prior is set as $Dir(N/K,\dots,\allowbreak N/K)$ for the $K$ potential regimes. When the $k$-th ($k$=$1,\dots,K-1$) regime finishes, suppose the observed epochs for past regimes $1,\dots,k$ are $m_1,\dots,m_{k-1}$, we update the posterior distribution to $Dir(m_1,\dots,m_{k}, S_k, \dots, S_k)$,  where $S_k=(N-\sum_{k=1}^K m_k)/(K-k)$. 
We compare the Bayesian update rule with the restatement rule in Figure~\ref{fig:dirichlet} and find the restatement rule has a lower interpolation error; the interpolation error is averaged over 200 jobs randomly drawn from the Gavel workload trace (Section \ref{subsec:expsetup}), each with a batch size scaling schedule imposed by Accordion or GNS.

\noindent\textbf{Predicting job remaining time. } Given the predictions from the Bayesian model, we next predict the remaining runtime for a job. This is necessary for estimating finish time fairness. We sum up individual regimes' expected duration to calculate total job runtime. Total job time minus cumulative run time in the past (i.e., $T_j$) gives the remaining time.  

\noindent\textbf{Computational tractability. }
 Finally, as each job can comprise of many possible regime trajectories, at the cluster level, the trajectory space for all jobs is combinatorially large. To avoid space explosion, the scheduler only considers a single regime transition trajectory for each job, which is the mean (expectation) of its posterior distribution model~\cite{barabaasi1999mean}.

\noindent \textbf{Evaluating prediction accuracy.} Figure~\ref{fig:dirichlet} shows the \underline{online} prediction (i.e., mean of posterior distribution) accuracy for regime transition and job run time. We compare \project's restatement rule with two baselines. The first is a standard Bayesian posterior update rule; the second is a greedy approach that forecasts future job run time only using the most up-to-date job throughput, which is used by all reactive schedulers.
The evaluation includes 200 Accordion and GNS jobs with real dynamic adaptation trajectories. \project's restatement rule converges to the oracle job run time and the oracle dynamic adaptation trajectory faster than the baselines. Throughout the training, the error in modeling the duration of each regime is on average 6\%, which results on average an 84\% accuracy in run time prediction.
In summary, we see that our proposed predictor for elastic training jobs is able to accurately capture the total run time without prior training and by only observing job progress across epochs. We next discuss how our predictor can be integrated with the market formulation.

\section{\project Design}
\begin{figure}[!t]
\includegraphics[width=\columnwidth]{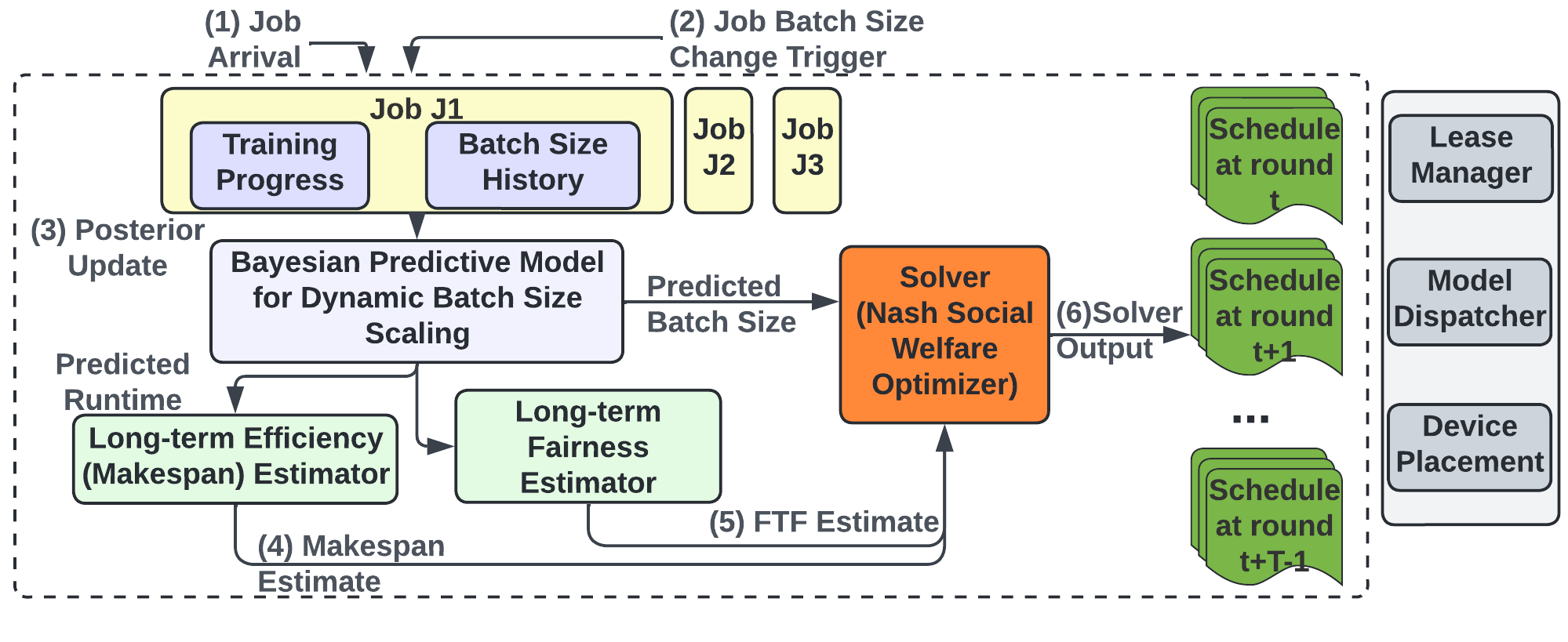}
\caption{Design of \project{} showing how the different components interact with each other to derive a schedule.} \label{fig:system_architecture}
\vspace{-0.1in}
\end{figure}

\noindent\textbf{Overview.} Figure~\ref{fig:system_architecture} presents the overall system design of \project. When a new job arrives (1), the Bayesian predictor will construct a prior model for the job's batch size scaling schedule and the job is added to the active pool. 

As a job makes progress, upon epoch completion or when the job triggers a dynamic batch size scaling (2), the job's Dirichlet posterior model is updated using the restatement rule (3). The posterior model then forecasts the future batch size schedule for this job and delivers it to the scheduler solver.
Further, the posterior model predicts the job's (remaining) run time under dynamic batch size scaling and delivers this to the long-term efficiency and fairness estimator. 

We design two estimators: a long-term efficiency estimator (4) that can estimate the makespan (time to finish all active jobs) and a long-term fairness estimator (5) that can estimate FTFs for all active jobs. The schedule solver converts the predicted batch size schedules into the utility for each job and synthesizes a generalized Nash social welfare function (6) that uses jobs' FTF estimate as weights and the makespan estimate as a regularizer. Finally, the output of the solver is a schedule for the next $T$ rounds, and this schedule is used by the cluster manager to launch jobs (more details in \cref{sec:impl}).

We next discuss some design details of how the generalized Nash social welfare function is derived from its inputs. We also present an overview of how the efficiency and fairness estimators work. We include a more detailed explanation in Appendix~\ref{apdx:shockwave_design}.

\subsection{Schedule Solver} 
\label{subsec:solver}
\textbf{Output.} The solver plans the schedule for a configurable number of future rounds $T$ (the default is 20 two-minute-long rounds). Thus, the output is a $N\times T$ binary matrix $\mathsf{X}$, where $N$ is the total number of active jobs available for scheduling. $\mathsf{X}[j,t]=1$ ($\mathsf{X}[j,t]=0$) represents scheduling (descheduling) job $J_j$ in round $t$. %

\noindent \textbf{Objective.} The inputs to the schedule solver include the batch size schedule for all jobs, which can be used to derive their utility (epoch progress) $\mathsf{UTIL}_{j}$, the estimated FTFs $\hat{\rho_{j}}$, and the estimated makespan $H$.

The objective of the solver is to maximize the generalized Nash social welfare as shown in Equation~\ref{eqn:schedule-solver-main}. $\sum_t \mathsf{UTIL}_j(\mathsf{X}[j,t])$ represents the summed utility of all active jobs. %
The utility increases when a job is scheduled for more rounds within the planning window, and the sum of the logarithm of utilities, across all jobs, represents the Nash social welfare. We use the k-th (default: 5) power of FTF values $\hat{\rho_{j}}$ as weights to prioritize jobs that are at risk of violating FTF (e.g., jobs that have been waiting in the queue for a long time). Finally, we add a regularization term that penalizes schedules (in the planning window) that could potentially increase the makespan estimate $H(X)$. Coefficient $\lambda$ (default: $1\mathrm{e}{-3}$) controls the magnitude of the regularizer, $Z_0$ is a normalization factor that renders the regularizer insensitive to the scale of  $H(X)$, and $M$ is the total number of GPUs in the cluster:

\begin{align}\label{eqn:schedule-solver-main}
\begin{split}
\underset{\mathsf{X}}{\mathsf{Maximize}}\; \frac{\sum_{j=1}^{N} {\widehat{\rho}(j)^{k}}log\,\sum_t \mathsf{UTIL}_j(\mathsf{X}[j,t])}{NM} -\frac{\lambda\mathsf{H(\mathsf{X})}}{Z_0}
\end{split}
\end{align}

We tune the hyperparameters over a large range and find that \project performs consistently well around the default hyperparameter values ($k$ in [1,10] and $\lambda$ in $[1\mathrm{e}{-4},  1\mathrm{e}{-2}]$). Exceedingly large or small hyperparameters make the regularization term dominate the Nash social welfare term (or vice versa) and push \project away from the Pareto frontier of fairness and efficiency, while the default values strike a balance between them.

Similar to prior work \cite{themis20, qiao2020pollux, gavel20}, the solver recomputes the program in Equation~\ref{eqn:schedule-solver-main} either when the planned rounds elapse, or when jobs arrive or complete. If dynamic adaptation is predicted to occur within the planning window, the scheduler needs to incorporate dynamic changes in jobs' throughputs when computing the utility. To account for dynamic changes, we decompose a job's schedule into regimes, where each regime has a fixed batch size and throughput. The generalized Nash social welfare (Equation~\ref{eqn:schedule-solver-main}) can then be implemented at a regime level, where the utility of a job equals the summed utility over all regimes.

\subsection{Long-term Fairness and Efficiency Estimators}  

\noindent\textbf{Finish time fairness estimator.} We estimate job $J_j$'s finish time fairness (FTF) $\widehat{\rho}(j)$ as its predicted job completion time (the sum of attained service time, waiting time, and the predicted remaining run time), divided by its predicted total job run time. Note that a job's predicted runtime is related to its predicted batch size scaling schedule. \project plugs in FTF $\rho$s of jobs into social welfare function (see Equation \ref{eqn:schedule-solver-main}) as weights. The weights in the social welfare function act as the budgets assigned to jobs in the volatile Fisher market. If a job is predicted to be unfairly scheduled (large FTF $\rho$) in long term, VFM correspondingly assigns a higher budget for it and proactively prioritizes the job in the planning window. 

\noindent\textbf{Makespan estimator.} The efficiency estimator estimates the makespan to complete all active jobs and penalizes schedules in the planning window that increase the makespan. However, it is challenging to estimate the makespan for all active jobs at a given instant. Thus, in practice, \project uses a lower bound~\cite{coffman1978application} of the makespan as a proxy and penalizes increasing the lower bound. More details are in Appendix~\ref{apdx:shockwave_design}.

\begin{table}[!tb]
\small
\setlength\extrarowheight{-3pt}
\begin{tabularx}{\columnwidth}{l l l l}
    \hline
    \makecell[l]{Model} & \makecell[l]{Task} & \makecell[l]{Dataset} & \makecell[l]{Batch Size(s)}\\ \hline
    \makecell[l]{ResNet-50} & \makecell[l]{Image\\ Classification} & \makecell[l]{ImageNet} & \makecell[l]{16 - 128}\\
    \hline
    \makecell[l]{ResNet-18} & \makecell[l]{Image\\ Classification} & \makecell[l]{CIFAR-10} & \makecell[l]{16 - 256}\\
    \hline
    \makecell[l]{LSTM} & \makecell[l]{Language\\ Modeling} & \makecell[l]{Wikitext-2} & \makecell[l]{5 - 80}\\
    \hline
    \makecell[l]{Transformer} & \makecell[l]{Language\\ Translation} & \makecell[l]{Multi30k\\ (DE-EN)} & \makecell[l]{16 - 256}\\
    \hline
    \makecell[l]{Recoder\\ Autoencoder} & \makecell[l]{Recommen-\\dation} & \makecell[l]{ML-20M} & \makecell[l]{512 - 8192}\\
    \hline
\end{tabularx}
\caption{\label{tab:workload}Workloads used in the evaluation.}
\end{table}

\section{Implementation}\label{sec:impl}
\noindent \textbf{Scheduler and worker.} \project scheduler and worker implement time-sharing of cluster resources with round-based scheduling. Each round is a fixed interval (default: 2 minutes). In each round, the scheduler selects a set of jobs from the active job pool to run. The lease manager translates the schedule to job leases and notifies the workers to launch, suspend, or resume jobs. Each worker binds to a single GPU device.

We adopt a simple job placement engine along with Gavel. The placement engine tries to tightly pack jobs' workers over the machines to minimize fragmentation, and it also tries to place scheduled jobs on their previously executed machines to maximize job locality.

\noindent \textbf{Scheduler solver, lease manager, and model dispatcher.} If a job does not run in round $T$, but is scheduled for round $T+1$, the scheduler will notify the lease manager to create a new lease for it, and dispatch the job to GPU workers before the next round starts. The assigned workers will launch the job when the next round begins. If a job is actively running in round $T$ and the scheduler continues to schedule it for round $T+1$, the lease manager will send a lease extension signal to the job's workers. This job will stay running on the same workers in round $T+1$. If a job is actively running in round $T$, but the scheduler decides to suspend it in round $T+1$, the job's workers will stop it since its lease will not be renewed. 

\project also penalizes frequent restarts as it adds overheads in dispatching models and datasets to workers. The schedule solver prefers to schedule jobs to continuous rounds in the window and penalizes scattering the job's execution across rounds. Furthermore, the underlying device placement engine prefers mapping a job to its previously allocated workers to reduce restarts.

\noindent \textbf{Dynamic adaptation support.} When a training job triggers dynamic adaption (i.e, batch size scaling), it notifies the scheduler solver of the occurrence of the event. The cluster manager can configure \project's responsiveness to dynamic scaling. The reactive mode requires \project to invalidate its current schedule and immediately trigger resolving in response to dynamic adaptation. The lazy mode continues the original schedule and postpones resolving until the next rescheduling interval. \project is by default configured in reactive mode.

\noindent \textbf{Prototype.} \project is implemented in Python atop ML cluster manager Gavel \cite{gavel20}. We integrate \project into Gavel by implementing a schedule solver, meta-data collector, and schedule translator, which translates \project's produced schedule to job leases. Furthermore, \project provides an interface for users to monitor gradients and trigger batch size scaling. Scaling requests are sent to the scheduler with gRPC. The schedule solver is implemented with Gurobi \cite{pedroso2011optimization}. \project uses Linux NFS to store model checkpoints. Our checkpointing overhead is less than 3\%.

\section{Evaluation}
We next evaluate \project using ML job traces derived from real-world clusters and compare \project to state-of-the-art deep learning schedulers. 
\subsection{Experiment Setup}\label{subsec:expsetup}
\textbf{Testbed.} We conduct experiments using a 32-GPU, 8-node cluster on TACC \cite{cazes2021preparing}. Each node has 4 NVIDIA Quadro RTX 5000 GPUs (16GB GRAM), 2 Intel Xeon E5-2620 v4 “Broadwell” CPUs, and 128GB DDR4 RAM. The network bandwidth is 200 GB/s inter-switch and 100 GB/s inter-node. %

\noindent\textbf{Workload.} \project’s evaluation uses two separate workloads to reinforce its practical applicability. These traces include diversity in job sizes, model types, and arrival patterns. Unless otherwise specified, we use Gavel's workload generator \cite{gavel20} to construct synthetic distributed training workloads. Job information is detailed in Table~\ref{tab:workload}. The jobs used in this paper range from 0.2 to 5 hours long, with 1, 2, 4, or 8 workers for distributed training, and the arrival of jobs follows a Poisson arrival process with an inter-arrival rate $\lambda$ ranging from 0.1 to 0.2 \cite{gavel20}.  We use a mix of job durations also derived from prior work~\cite{qiao2020pollux}. We categorized jobs based on total GPU-time, and similar to prior work, we set the probability of generating Small (0.2-8 GPU-hours), Medium (8-16 GPU-hours), Large (16-72 GPU-hours), and Extra Large (>72 GPU-hours) jobs to be 0.72, 0.2, 0.05, 0.03, respectively. Each job is configured with one of the three modes: Static, Accordion \cite{agarwal2021adaptive}, or GNS \cite{gns_scaling}. We increase the total batch size by increasing the per-GPU batch size while preserving the number of workers. 
In addition to traces generated by Gavel, we also evaluate a production trace of real job duration and arrival timestamps used by Pollux~\cite{qiao2020pollux} in Appendix \ref{subsec:eval_pollux_arrival}.
We also tune the hyperparameters $k$ and $\lambda$ with the range discussed in Section \ref{subsec:solver}.

\subsection{Baseline Schedulers}\label{subsec:baselines}
We compare \project to six schedulers: \textbf{OSSP (Open Shop Scheduling) \cite{gonzalez1976open}, AlloX~\cite{allox19}, Themis \cite{themis20}, Gavel \cite{gavel20}, MSS (Max-Sum-Throughput) \cite{gavel20}, Gandiva-Fair \cite{chaudhary2020balancing}, and Pollux \cite{qiao2020pollux}}. All baselines, except Pollux, do not change the number of workers, whereas Pollux dynamically tunes the number of workers (and batch size) to adapt to varied resource availability.
To perform a fair comparison against the scheduling policies of most baselines, in our \project{} prototype, we only perform time-sharing and maintain a fixed number of workers through a job's lifetime, even though the \project market formulation can be easily re-parameterized to support worker scaling. Nevertheless, we compare our "constrained" version of \project to Pollux in \cref{subsec:pollux_eval} and show significant fairness gains and matching efficiency.

\textbf{Efficiency baseline: makespan.} OSSP minimizes makespan using MILP. As minimizing makespan usually translates to maximizing cluster utilization, OSSP provides a baseline for efficiency, but with no guarantee of fairness. 

\textbf{Efficiency baseline: throughput.} Max-Sum-Throughput (MST) maximizes the cluster-level throughput at each instant, which is the sum of throughput across all training jobs. %
MST is an instantaneous efficiency baseline.

\textbf{Fairness baseline.} Gavel~\cite{gavel20} implements Max-Min-Fairness~\cite{nace2008max}, an algorithm that performs fair sharing of cluster resources within each allocation round. 

\textbf{Fairness and responsiveness baseline.} AlloX \cite{allox19} minimizes average job completion time with maximal bipartite matching and provides a baseline for responsiveness. 
Pollux \cite{qiao2020pollux} maximizes cluster-wide goodput and uses the $p-norm$ of individual jobs' training goodput for improved responsiveness, while tuning $p$ to penalize unfair allocations.

\textbf{Fairness and efficiency baseline.} Themis \cite{themis20} uses Partial Allocation\cite{partalloc13} for efficient and fair allocation. We use the default filter value for Themis. We also compare against Gandiva-Fair~\cite{chaudhary2020balancing}, a framework that uses lottery scheduling to guarantee a proportionally fair share of resources and efficiency by being work-conserving.

\noindent\textbf{Performance metrics.} We quantify efficiency using makespan and utilization. We measure fairness using two metrics: The first is the fraction of unfairly scheduled jobs, i.e., the fraction of jobs with FTF $\rho>1.0$; The second is the worst-case FTF $\rho$, which is the worst-case slowdown due to unfair scheduling. The smaller the unfair fraction and worst FTF $\rho$ are, the better a scheduler is at preserving sharing incentive. We quantify responsiveness using average JCT.

\begin{figure}[!tbp]
\centering
\includegraphics[width=0.5\textwidth, trim=0.1cm 0.1cm 0.1cm 0.1cm,clip]{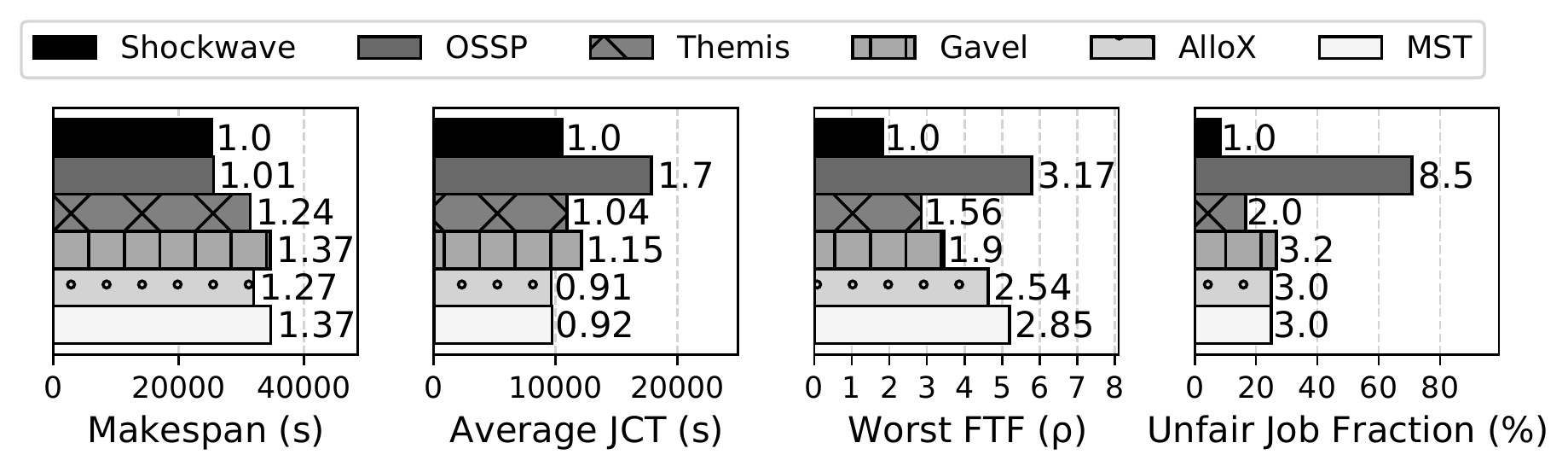} 
\caption{[Physical] Evaluating \project's scheduling efficiency and fairness in a 32-GPU physical cluster. The annotated number beside each bar is the relative value compared to \project.}
\vspace{-0.1in}
\label{fig:physical-tacc}
\end{figure}

\subsection{Evaluating Efficiency and Fairness}
We first study the benefits of \project{} using experiments on the physical TACC cluster.

\begin{figure*}[!tbp]
\centering
\begin{subfigure}[b]{0.64\textwidth}
    \includegraphics[trim=0.05cm 0.1cm 0.1cm 0.1cm, clip, width=\textwidth]{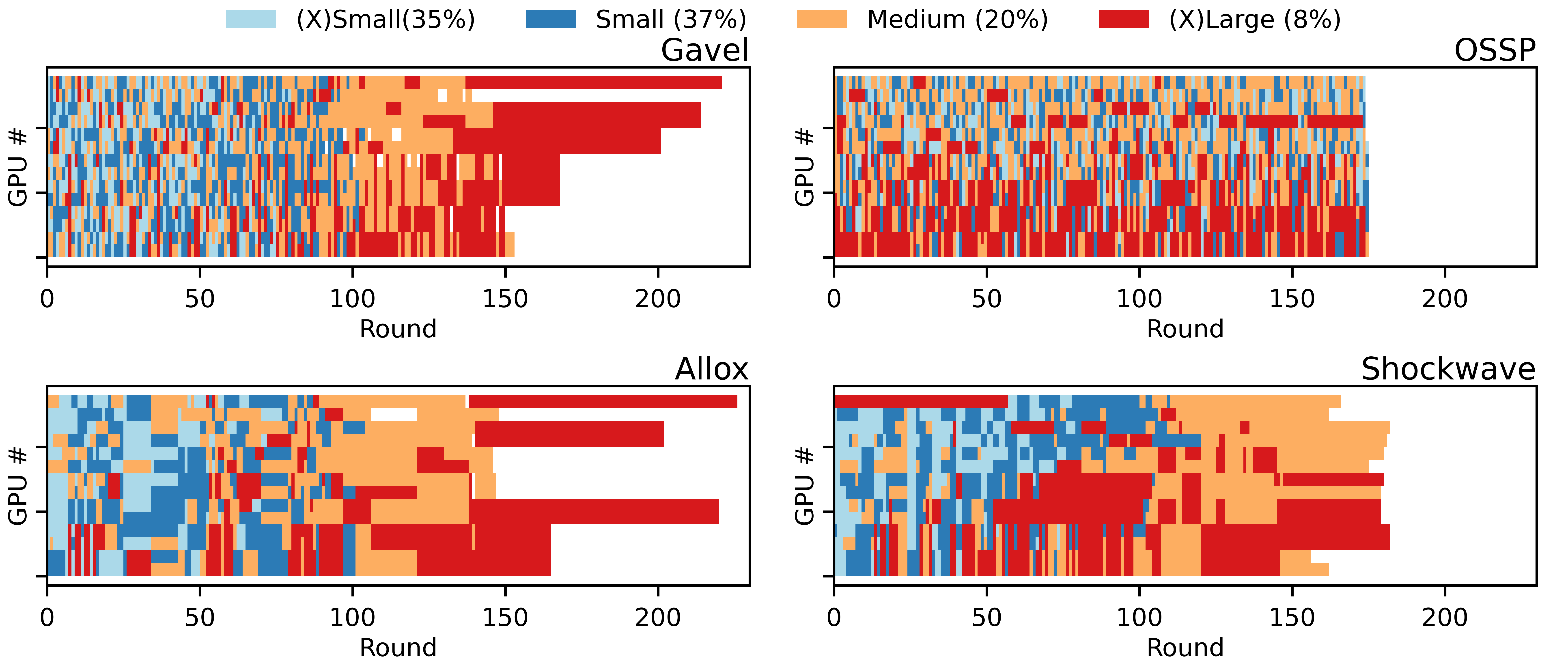}
    \vspace{-0.2in}
    \caption{Visualized schedules.}\label{fig:schedvis-sched}
\end{subfigure}
\hspace{0.2in}
\begin{subfigure}[b]{0.2\textwidth}
    \includegraphics[trim=0.1cm 0.1cm 0.1cm 0.1cm, clip, width=\textwidth]{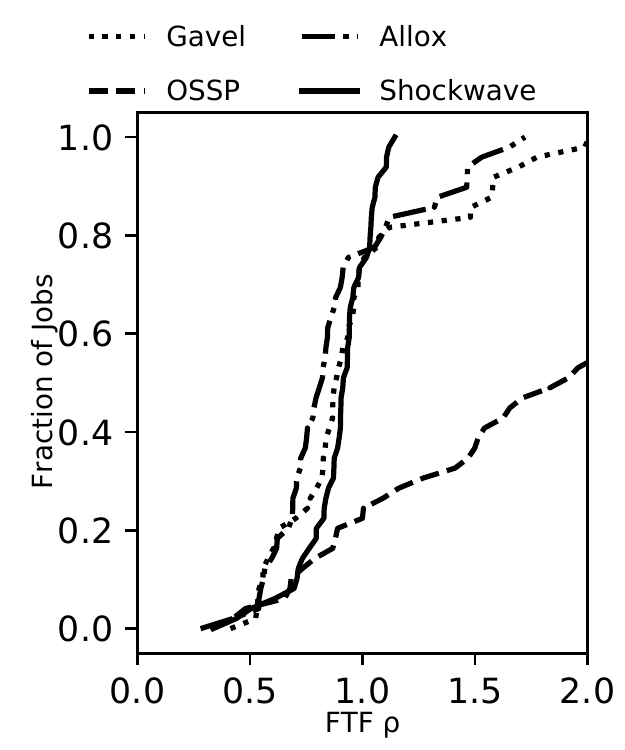}
    \vspace{-0.2in}
    \caption{FTF $\rho$ CDF}\label{fig:schedvis-ftfs}
\end{subfigure}
\caption{A closer look at how \project prioritizes jobs of different lengths while meeting the fairness constraints.}\label{fig:schedvis}
\vspace{-0.15in}
\end{figure*}

\textbf{[Cluster - 32 GPUs, 120 Jobs] Efficiency. (cf., Figure~\ref{fig:physical-tacc})} \project is more efficient than existing fair schedulers with a makespan on average 1.3$\times$ less than Themis, Gavel, and AlloX.
Compared to our efficiency baselines that have no fairness constraints,
\project achieves a 37\% improvement in makespan over MST and produces a similar makespan as OSSP. Analyzing cluster utilization data we also find that \project outperforms Themis, Gavel, and AlloX in cluster utilization by 28\% on average.

\textbf{[Cluster - 32 GPUs, 120 Jobs] Finish time fairness (cf., Figure~\ref{fig:physical-tacc}).} \project is fairer than existing fair schedulers. \project's worst-case FTF (Finish Time Fairness) $\rho$ is 1.82, outperforming Themis, Gavel, and AlloX by 2$\times$ on average. OSSP and MST are not fair schedulers, and severely break finish time fairness, the worst-case FTF $\rho$ of which reach 5.79 and 5.2. In addition, \project keeps the fraction of unfairly scheduled jobs (i.e., the fraction of jobs with FTF $\rho$>1) low, outperforming Themis, Gavel, and AlloX by 2.7$\times$ on average. OSSP and MST unfairly schedule jobs, and their fraction of jobs that have FTF $rho$ larger than 1 are 70.8\% and 25\%.

\textbf{[Cluster - 32 GPUs, 120 Jobs] Average job completion time (cf., Figure~\ref{fig:physical-tacc}).} \project does not sacrifice system responsiveness in exchange for improved makespan and finish time fairness. \project produces a similar average job completion time when compared with Themis, Gavel, and MST. AlloX achieves a better average JCT by aggressively prioritizing short jobs (but at the cost of delaying long jobs), while in contrast, OSSP achieves the worst average JCT due to aggressively prioritizing long jobs for tight resource packing over time (but at the cost of delaying short jobs).

Overall, we find that by solving for the optimal efficiency-fairness trade-off, \project can improve efficiency and fairness when compared with existing schedulers. By analyzing the scheduling decisions, we find that with \project{}, jobs are opportunistically prioritized to improve long-term efficiency if such prioritization does not affect finish time fairness. Second, we find that \project's solver improves fairness by smart arbitrating. ``Rich'' jobs (i.e., jobs which have lower chances of violating FTF) yield resources to ``poor'' jobs which have a higher chance of violating FTF. We next take a closer look at the schedule decisions on a smaller trace to further distill the benefits of \project{}.

\subsection{A Closer Look at \project's Schedule}\label{subsec:closer_look}
We compare the schedules for a batch of 50 jobs and the FTF $\rho$ between \project and baselines to further understand the wins in efficiency and fairness. We categorize jobs into four groups based on their sizes (GPU-time): (X)Large, Medium, Small, and (X)Small (different colors in Figure~\ref{fig:schedvis-sched}).

\noindent \textbf{Understanding efficiency improvement.}
\textbf{AlloX} optimizes system responsiveness (average JCT) by prioritizing small jobs. In Figure~\ref{fig:schedvis-sched}, in the first 100 rounds, most of the jobs scheduled are XSmall jobs. The filter in AlloX ensures medium and large jobs do not get starved but these jobs are not prioritized. Large jobs trail until round 230 and leading to worse makespan and cluster utilization.

\textbf{Gavel}'s max-min fair scheduling prioritizes the least performant jobs. In Figure \ref{fig:schedvis-sched}, throughout the schedule, Gavel prioritizes neither small nor large jobs; jobs of all sizes evenly partition GPUs when compared with other policies. However, restricting scheduling to instantaneous fairness significantly hurts long-term efficiency, and large jobs run on a mostly idle cluster from round 170 to round 220. 

\textbf{\project} improves efficiency by opportunistically scheduling (X)Large jobs across rounds but without hurting small and medium jobs' sharing incentive (Figure\ref{fig:schedvis-ftfs} shows the CDF of FTF).
In Figure~\ref{fig:schedvis-sched}, between round 0 and 50, and 50 to 120, large jobs are opportunistically scheduled by \project, and the cluster is tightly packed over the time horizon, resulting in a low makespan. Note that \project also preserves responsiveness since most XSmall and small jobs are completed fast (before round 50 and 110, respectively), which is comparable to AlloX.

\textbf{OSSP} over-prioritizes (X)Large and medium jobs throughout the timeline, but significantly delays XSmall jobs' completion (see delayed blocks at the end of the schedule). Delaying small jobs significantly breaks the sharing incentive and undermines cluster responsiveness.

\noindent \textbf{Understanding fairness improvement.} 
Figure \ref{fig:schedvis-ftfs} shows the FTF $\rho$ CDFs of different policies for the batch of jobs visualized in Figure~\ref{fig:schedvis-sched}. \project improves efficiency without sacrificing the sharing incentive: the worst-case FTF $\rho$ for the batch of jobs is 1.23, and the fraction of unfair jobs is low. 
In Figure~\ref{fig:schedvis-sched}, AlloX and Gavel's CDF grows faster than \project's for $\rho<=1$, although more than 20\% of jobs have $\rho>1$. AlloX and Gavel over-prioritize some jobs and this results in an allocation that exceeds sharing incentive. \project avoids over-prioritization and is thus able to have more jobs meet the sharing incentive. \project also improves fairness by predicting dynamic adaptation for a more accurate estimate of the FTF deadline. Figure~\ref{fig:motivation-ftf-adapation} shows an example where \project produces an accurate prediction of the FTF deadline and enables the job to finish on time.

\subsection{Scaling to Large Clusters}
We next use simulation to compare \project's and baseline algorithms' efficiency and fairness in larger-scale cluster settings. 
We scale both the cluster size and the number of jobs and study 64 GPUs with over 220 jobs, 128 GPUs with over 460 jobs, and 256 GPUs with over 900 jobs. We preserve the contention factor as roughly three to maintain a constant level of resource contention regardless of scale. Note that our physical cluster implementation and the simulator use the same scheduling code base and solver engine. We begin by validating our simulator's fidelity.

\noindent\textbf{Simulation Fidelity}\label{subsec:simfidelity}

We evaluate the simulation fidelity by comparing our simulator's results with the 32 GPU physical cluster results (Table~\ref{tab:fidelity}). We run all policies supported by our system under different workloads, and the average difference is reported in Table~\ref{tab:fidelity}. Overall, the performance difference between a simulated and physical cluster run is around 5\%.

\begin{table}[!tbp]
\small
\center
\begin{tabular}{|c|c|c|}
    \hline
    Makespan (s) & Average JCT (s) & Unfair Fraction (\%) \\ 
    \hline
    4.97\% & 4.62\% & 3.83\% \\
    \hline
\end{tabular}
\caption{\label{tab:fidelity} Fidelity of \project's simulator -- difference between simulator and physical cluster.}
\vspace{-0.1in}
\end{table}

\begin{figure}[!tbp]
 \centering
\includegraphics[width=0.48\textwidth, trim=0.1cm 0.1cm 0.1cm 0.1cm,clip]{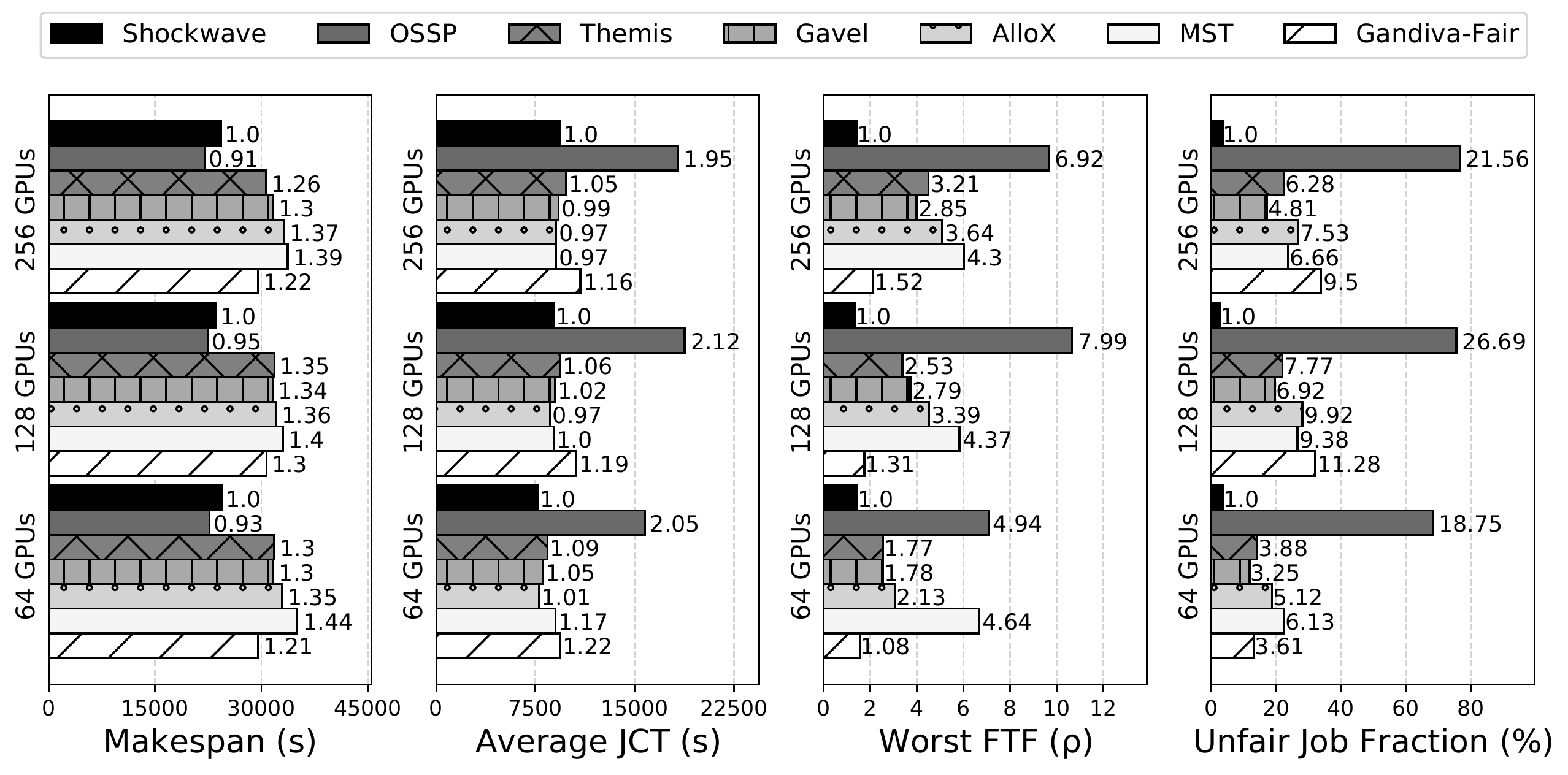} 
\caption{[Simulation] Evaluating \project's scheduling efficiency and fairness in differently sized large clusters.}
\label{fig:large-cluster-sim}
\vspace{-0.1in}
\end{figure}

\noindent\textbf{[Simulation - 64-256 GPUs, 220-900 Jobs] Efficiency.} As shown in Figure~\ref{fig:large-cluster-sim}, \project scales to large cluster settings and preserves the improvement in makespan over baseline algorithms. \project achieves 1.26-1.35$\times$, 1.3-1.34$\times$, 1.35-1.37$\times$, and 1.21-1.3$\times$ speedup in makespan when compared with Themis, Gavel, AlloX, and Gandiva-Fair respectively. \project achieves a marginally worse (5\%-9\%) makespan compared with OSSP.

\noindent\textbf{[Simulation - 64-256 GPUs, 220-900 Jobs] Finish time fairness.} The worst-case FTF $\rho$ for \project when scaling to large clusters is on average 1.32, outperforming fair scheduling policies Themis, Gavel, AlloX, and Gandiva-Fair by 2.5$\times$, 2.4$\times$, 3.1$\times$, and 3.9$\times$ respectively. In addition, \project maintains the fraction of unfairly scheduled jobs (FTF $\rho$ > 1) on average at 4\%, outperforming other fair scheduling baselines by 6$\times$.

\noindent\textbf{[Simulation - 64-256 GPUs, 220-900 Jobs] Average job completion time.} At a large scale, \project maintains similar responsiveness when compared with fair schedulers. One exception here is Gandiva-Fair which prolongs average JCT by 16-22\%. Gandiva-Fair uses stride scheduling~\cite{waldspurger1995lottery} where, by default, a job's number of tickets is equal to the job size (i.e., the number of workers). Thus, large jobs have a higher proportional share when compared with small jobs, and can delay small jobs, thereby degrading system responsiveness.

We study the solver overhead with large clusters in \cref{subsec:solver-overhead}.

\begin{figure}[!tbp]
  \centering
  \includegraphics[width=0.5\textwidth, trim=0.1cm 0.1cm 0.1cm 0.1cm,clip]{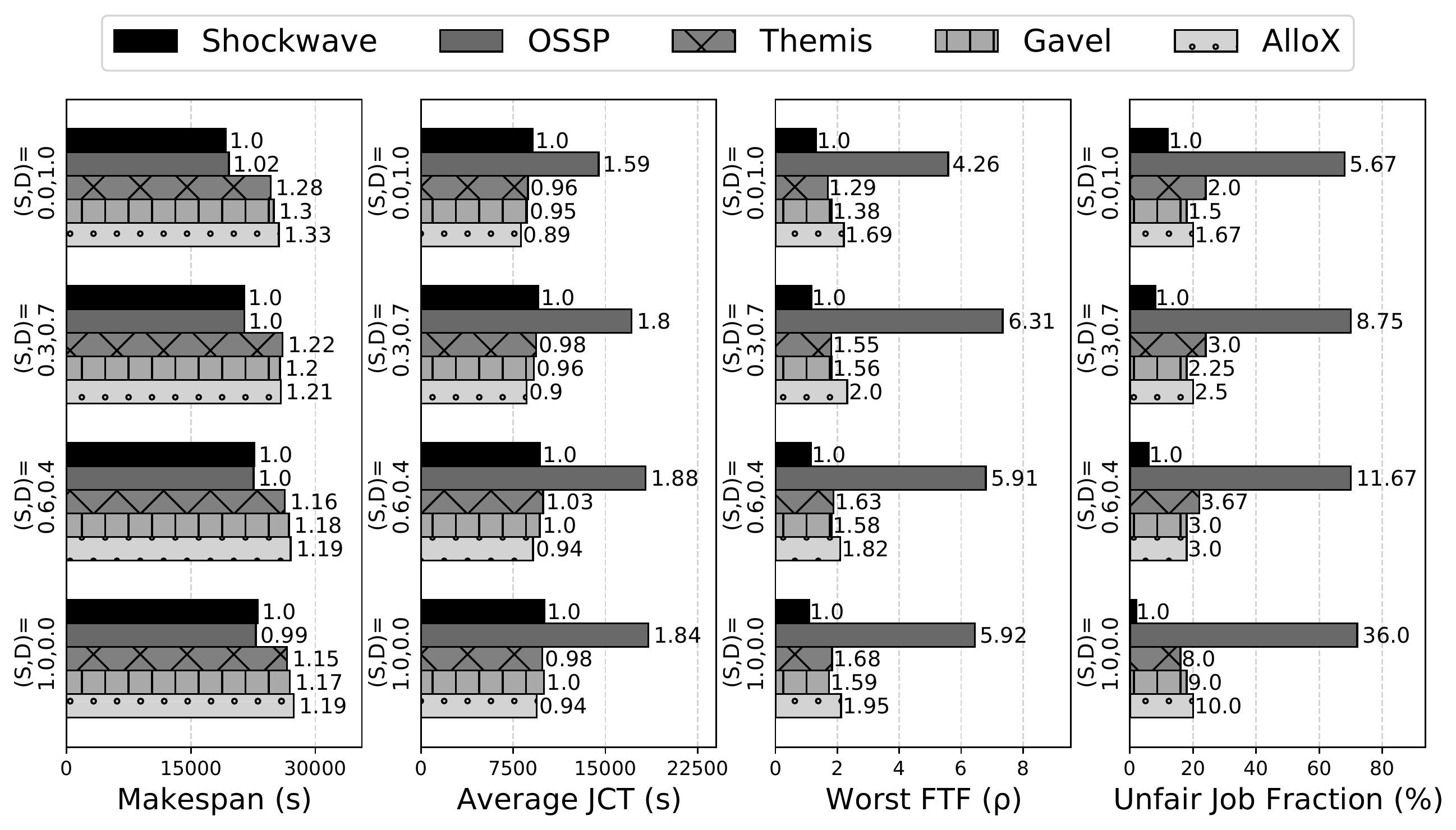}
  \caption{[Simulation] Effects of varying the mix of static and dynamic Jobs. (S, D)=(x, y) indicates x fraction of static jobs and y fraction of dynamic jobs.} \label{fig:mode-mix}
  \vspace{-0.1in}
\end{figure}%

\subsection{Benefits of Proactive Scheduling}\label{subsec:proactive}
We next compare \project and baseline policies while varying the mix of static and dynamic jobs in simulation.

\noindent \textbf{All static jobs.} We first analyze the case where all jobs disable dynamic adaptation. This isolates \project's win due to social welfare maximization. The results (Figure~\ref{fig:mode-mix}) show that all fair scheduling policies, i.e., \project, Themis, Gavel, and AlloX exhibit a relatively low fraction (<18\%) of unfairly scheduled jobs (FTF $\rho$>1.0), but \project outperforms the baseline algorithms by limiting the unfair fraction to less than 5\%. 
\project has on average an 18\% improvement in makespan over Themis, Gavel, and AlloX, with no loss in average JCT. Overall, these results show how maximizing social welfare over time can achieve a better fairness-efficiency trade-off when compared to existing approaches.

\noindent\textbf{Fairness and efficiency while being proactive.} \project sees a larger win in makespan as the fraction of dynamic jobs increases. The speedup over Gavel, Themis, and AlloX increases to 1.3$\times$ when the fraction of dynamic jobs grows from 0.4 to 1.0.  Our results also show that existing schedulers that are reactive to dynamic scaling have suboptimal fairness outcomes. Both Themis and AlloX exhibit an increased unfair job fraction as the number of dynamic jobs increases.  When all jobs are dynamic, Themis schedules 28\% of jobs unfairly and AlloX schedules 22\% of jobs unfairly, while \project has a relatively (9\%) low fraction of unfairly-scheduled jobs. %

\subsection{\project versus Pollux}\label{subsec:pollux_eval}

\begin{figure}[!tbp]
 \centering
\includegraphics[width=0.5\textwidth, trim=0.1cm 0.1cm 0.1cm 0.1cm,clip]{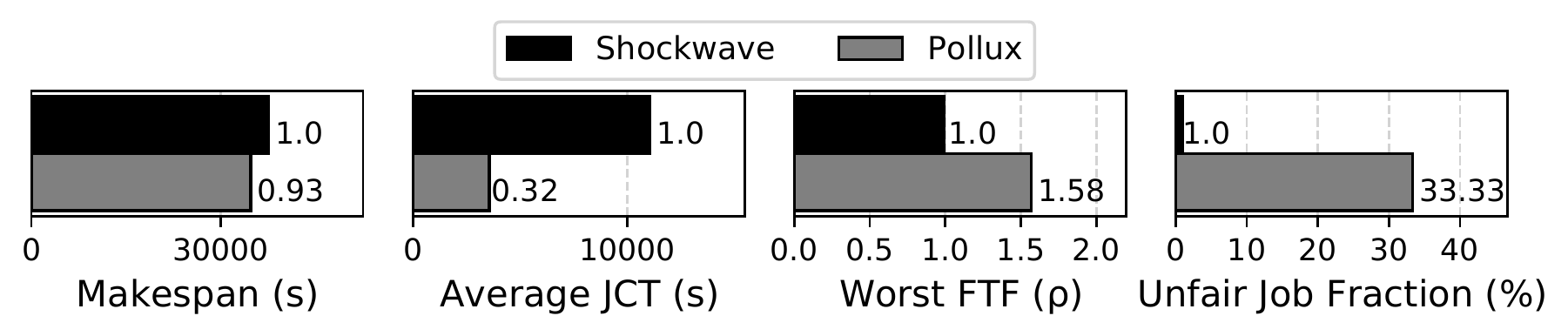} 
\caption{[Simulation] Evaluating \project's and Pollux's efficiency and fairness.}
\label{fig:shockwave-vs-pollux}
\vspace{-0.1in}
\end{figure}

To compare the scheduling policies used by Pollux and \project{}, we run both systems using the same workload trace provided by Pollux. We also first run the Pollux simulator to collect the batch size schedule observed at runtime and use that as an input to the \project simulator. Thus, both systems see the same set of input jobs and the same batch size schedule, and hence, job processing times should match even with dynamic scaling.

\textbf{JCT.} From Figure~\ref{fig:shockwave-vs-pollux}, we see that Pollux has a $3\times$ improvement in average JCT over \project. Pollux can scale the number of workers of a job, which leads to reduced resource contention and improved responsiveness. In fact, we found that Pollux reduces the requested GPU hours per job by $2.4\times$ when compared to the original trace. 
As our \project prototype does not change the number of workers used by a job, it preserves the contention level in the trace ($2.4\times$ larger than Pollux) and thus exhibits inferior responsiveness. We note that as seen in Figure~\ref{fig:physical-tacc}, \project has comparable JCTs with other baselines and the Pollux paper~\cite{qiao2020pollux} also reports a $3\times$ speedup over the baselines.

\textbf{Finish time fairness.} \project significantly outperforms Pollux w.r.t finish time fairness. This is because Pollux focuses on instantaneous fairness at each allocation but does not systematically address long-term fairness. At every round, Pollux's $p-norm$ formulation penalizes unfair allocations that lead to low instantaneous throughput for jobs but does not preserve long-term fairness over multiple allocation rounds. On the other hand, \project's dynamic market formulation provably guarantees long-term fairness.

\textbf {Makespan.} \project benefits from optimizing for long-term efficiency and has a similar makespan as Pollux despite not changing the number of workers dynamically.

Finally, we note that as discussed in~\ref{subsec:support_user_customization_bs} and Appendix~\ref{apdx:accuracy_gap_ncf}, Pollux's approach of automatically tuning the batch size and the number of workers can lead to accuracy loss (e.g., 2\% for ResNet18 and up to $4\%$ for DeepSpeech~\cite{qiao2020pollux}). We argue that Pollux's accuracy loss and poor fairness properties make it less attractive for practical deployments.

\subsection{Varying Cluster Contention and Workload}
We also vary the workload contention factor and compare all policies on a smaller 14-GPU physical cluster. We find that \project's fairness and efficiency win over the baseline schedulers increases (decreases) as cluster contention grows (drops). We include more details in Appendix~\ref{subsec:vary_contention}. We also compared \project using arrival patterns from the Pollux~\cite{qiao2020pollux} trace. Appendix~\ref{subsec:eval_pollux_arrival} includes these results.

\subsection{Solver Overhead}\label{subsec:solver-overhead}
\begin{figure}[!tb]
\centering
  \includegraphics[width=0.8\columnwidth]{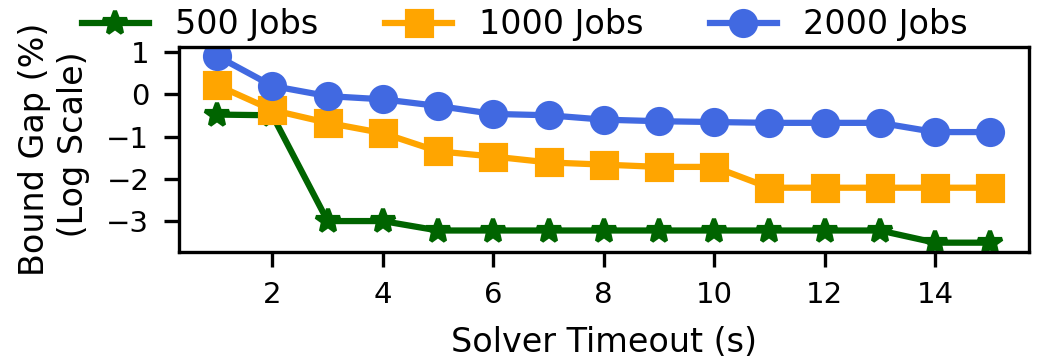}
  \caption{Solver Overhead. } \label{fig:overhead}
  \vspace{-0.1in}
\end{figure}
\project uses a timeout knob (default 15s) to limit the overhead of solving our market formulation. 
Figure~\ref{fig:overhead} uses simulation to show that on a 256 GPU cluster, the solver quality improves with diminishing returns as we increase the solver timeout from 1 second to 15 seconds. We measure solver quality using the \emph{bound gap} (how far the solution found at the timeout is from the optimal). The relative bound gap at 15 seconds is small (0.03\%, and 0.11\%) for 500 and 1000 active jobs. The bound gap at 15 seconds for 2000 jobs increases to 0.44\%. While this exceeds the criterion (0.1\%) recommended by Gurobi \cite{pedroso2011optimization}, our results show a limited impact on efficiency and fairness. We note that our solver runs in a separate thread and is proactively invoked in the middle of the current round. Thus, the solver overhead is \emph{hidden} when it is less than half-round duration.

\subsection{Resilience to Prediction Error}\label{subsec:solver-error}
\begin{figure}[!tb]
\centering
  \includegraphics[width=\columnwidth]{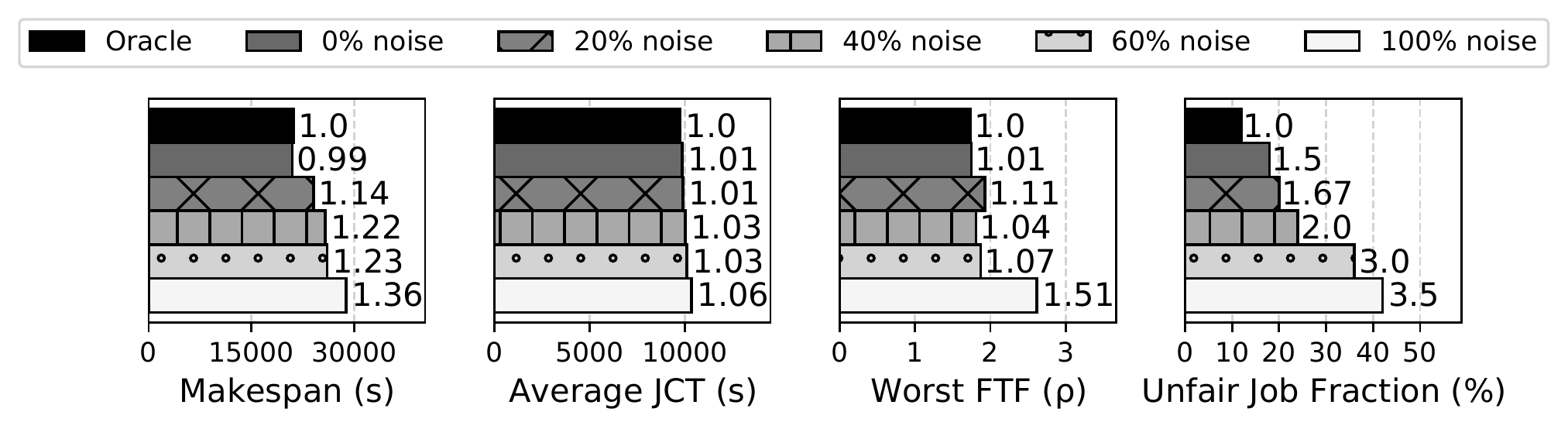}
  \caption{\project's scheduling efficiency and fairness under different levels of prediction errors.} \label{fig:rubustness}
  \vspace{-0.1in}
\end{figure}
Figure \ref{fig:rubustness} shows \project's resilience to prediction errors when varying levels of random noises (i.e., $\pm$ p\%) are injected into its interpolated job run time (under dynamic adaptation). The experiment settings in Figure \ref{fig:rubustness} are similar to those in Figure~\ref{fig:mode-mix} with all jobs enabled for dynamic batch size scaling (i.e., (S, D)=(0, 1.0)). First, we observe that as the injected errors grow, \project's worst-case FTF $\rho$ and the fraction of unfairly scheduled jobs inflate slowly. A similar steady trend holds for \project's average JCT. We argue such robustness originates from the design principle of Nash social welfare, which emphasizes common ownership and fair sharing of cluster resources; the penalty is huge if skewed training progress is present in the cluster and it leads the scheduler to be conservative to jobs' interpolated schedule slacks that are predicated by the biased FTF estimates. Second, we find that \project's scheduling efficiency drops as the errors grow. \project opportunistically prioritizes long-running jobs over the short ones to improve makespan. Having 100\% injected noise affects \project's estimation of job length and lowers its scheduling efficiency by over 30\%. Note that this deteriorated efficiency is still on par with the baseline schedulers (e.g., Themis, Gavel, and AlloX in Figure ~\ref{fig:mode-mix}).
\color{black}

\section{Related Work}
\label{sec:related}

We detail the comparison between \project and existing schedulers (e.g., Gandiva \cite{gandiva18}, Optimus \cite{optimus18}, DRF \cite{drf11}, REF \cite{ref2014}, Themis \cite{themis20}, AlloX \cite{allox19}, Tiresia \cite{gu2019tiresias}, Gandivar-Fair \cite{chaudhary2020balancing}) in Section \ref{sec:motivation}, and spotlight \project's contribution from two angles. First, \project is built on Nash social welfare, a theoretically-grounded approach to co-optimize long-term, rather than instantaneous, fairness and efficiency. Second, \project proactively plans schedules for dynamic adaptation, while most existing schedulers only react to dynamic adaptation. Section~\ref{sec:motivation} presented more details on the limitations of existing DL cluster schedulers.

AFS (Apathetic Future Share) \cite{hwang2021elastic} is another elastic sharing mechanism proactive to system dynamics. However, dynamic changes in AFS refer to job arrival and time-variant cluster contention, while jobs themselves {\em do not} change. \project has a different focus: jobs' resource demands (and efficiency) dynamically change due to batch size scaling. Further, AFS primarily focuses on improving average JCT while \project maximizes social welfare over time.

\section{Conclusion}
We presented \project, a market-theory-based efficient and fair scheduling framework for DNN training workloads. We showed how existing schedulers fail to preserve fairness and degrade efficiency by being reactive to dynamic adaption. To address these challenges, we proposed a proactive approach that uses dynamic markets and Bayesian statistics for scheduling. Our experiments show that \project can improve efficiency and fairness compared to state-of-the-art schedulers.

\paragraph{Acknowledgements:} We would like to thank the anonymous reviewers and our shepherd Zhihao Jia for their constructive comments that helped improve our paper. We would also like to thank Zhao Zhang for helping us run experiments on TACC resources and Mosharaf Chowdhury for feedback on an earlier draft of this paper. This work was supported in part by a University of Wisconsin Fall Research Competition grant,
by NSF grants CNS-2106199 and CNS-2105890 and by the CIFellows program, organized by the Computing Research Association and Computing Community Consortium. 

\bibliographystyle{acm}
\bibliography{bib}
\clearpage
\appendix
\section{Dynamic Batch Scaling Degrades Accuracy}\label{apdx:accuracy_gap}

\subsection{When does batch size scaling degrade accuracy}
Improper batch size scaling can adversely affect convergence and degrade the accuracy of the trained model. This is known as \textbf{generalization gap} \cite{hoffer17trainlonger, keskar2017large} and its underlying reasons are still not well understood. We next list different analyses on when batch size scaling can affect final model quality: (a) scaling up the batch size by $k\times$ reduces the number of iterations per epoch by $k\times$, and given a pre-specified number of epochs, it reduces the overall iterations of back-propagation by $k\times$. Accuracy loss stems from a reduced number of model updates~\cite{hoffer17trainlonger}. (b) Scaling up the batch size reduces the noise in the gradient estimate, but noise serves to regularize training and can navigate the optimizer away from local minima. Batch size scaling thus hurts generalization by reducing healthy gradient noises. (c) Scaling up the batch size causes training to converge to sharp minima, and the model outputs are sensitive to small perturbations in the input. This results in a poorer generalization~\cite{keskar2017large}. 

Researchers have developed heuristics \cite{increasebs18, agarwal2021adaptive} and adaptive batch size scaling techniques (e.g., Gradient Norm~\cite{agarwal2021adaptive}, GNS (Gradient Noise Scale)~\cite{gns_scaling} and Heissan Eigenspectrum~\cite{yao2018hessian}) to mitigate generalization gap, but no single technique handles all models, datasets and optimizers. Thus, today, there are many different batch size scaling techniques in the ML community. Pollux adopts GNS while recent work points out some of the limitations in applying GNS~\cite{qin2021simigrad} for batch size scaling.

\subsection{Example: Pollux's automatic batch size scaling leads to accuracy loss in NeuMF-m1-lm training}\label{apdx:accuracy_gap_ncf}
\begin{figure}[h]
   \centering
   \includegraphics[width=1.0\linewidth]{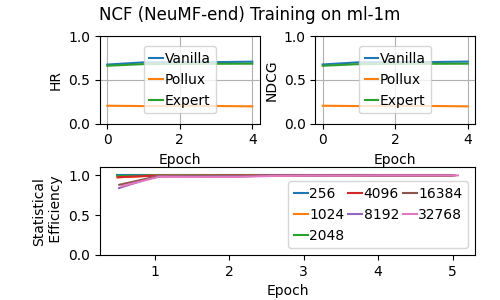}
   \caption{Comparing model accuracy (NCF-ml-1m) for vanilla training (no batch size scaling), expert-set batch size scaling, and Pollux's autoscaling. The legends in the bottom figure indicate batch size.}\label{fig:mv-example-accuracy_gap_ncf}
 \end{figure}
Figure~\ref{fig:mv-example-accuracy_gap_ncf} shows that statistical efficiency minimally degrades when scaling up from a batch size of 256 to 32768, and that this is true even for early training epochs. Therefore, Pollux immediately scales up the batch size from 256 to 32768 at epoch 1. However, we found that such early, aggressive scaling leads to inferior validation accuracy, i.e. lower HR (Hit Rate) and NDCG (Normalized Discounted Cumulative Gain), when compared with vanilla training where dynamic batch size scaling is disabled. An expert-set dynamic scaling schedule that scales up the batch size to 32768 at epoch 3 and this helps match the validation accuracy of vanilla training.
\section{Static Filters Degrade Efficiency, Fairness}\label{apdx:toy_example}

\vspace{-5mm}

\begin{figure}[htp]
    \subfloat[Themis with $f=1/3$.]{%
      \includegraphics[width=\columnwidth]{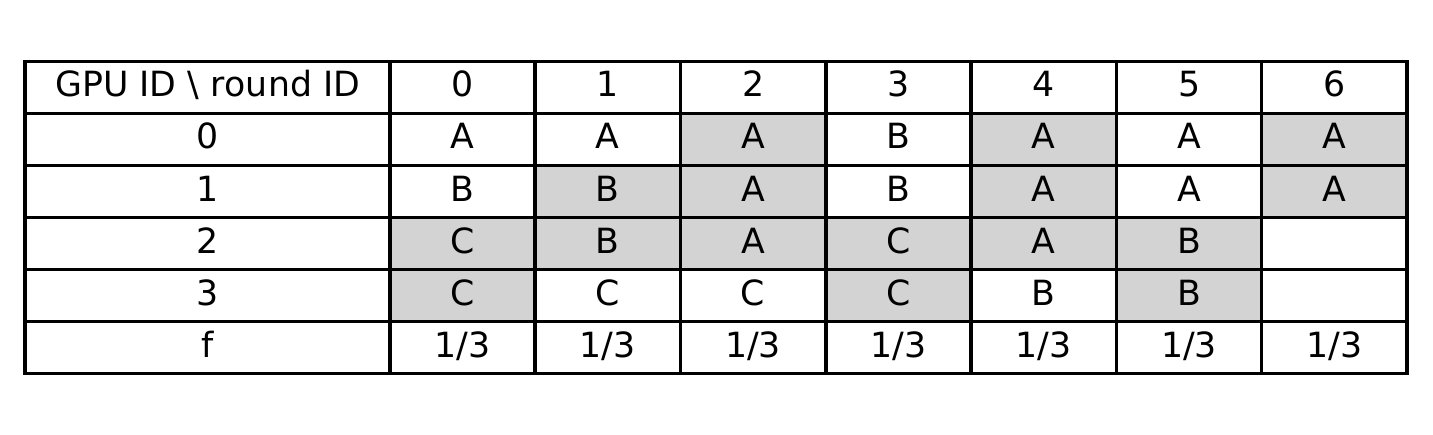}%
      \vspace{-5mm}
    }
    
    \subfloat[Themis with $f=1.0$.]{%
      \includegraphics[width=\columnwidth]{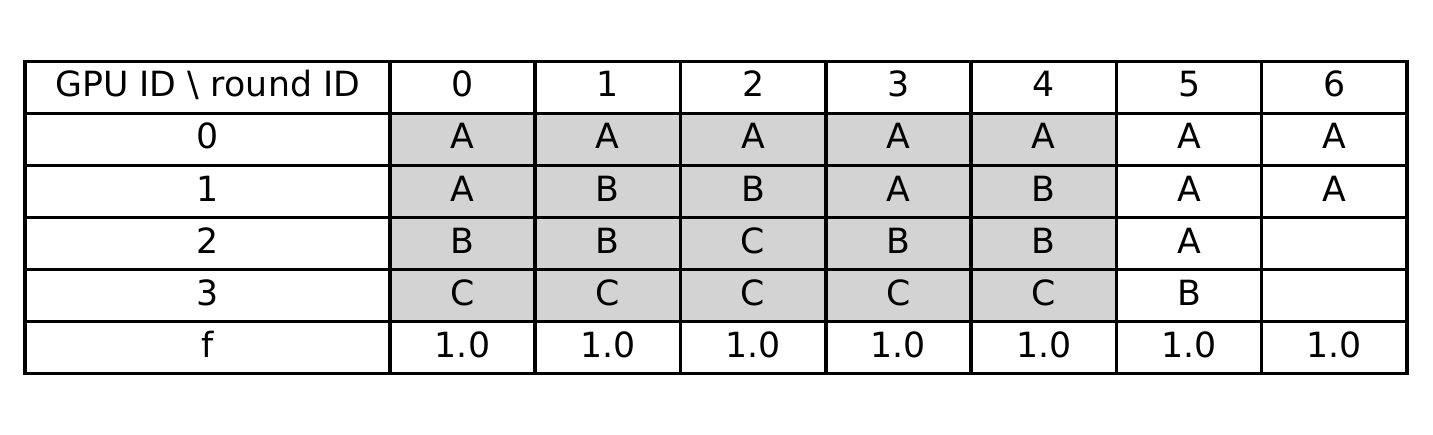}%
      \vspace{-5mm}
    }
    
    \subfloat[\project with a dynamic filter $f$.]{%
      \includegraphics[width=\columnwidth]{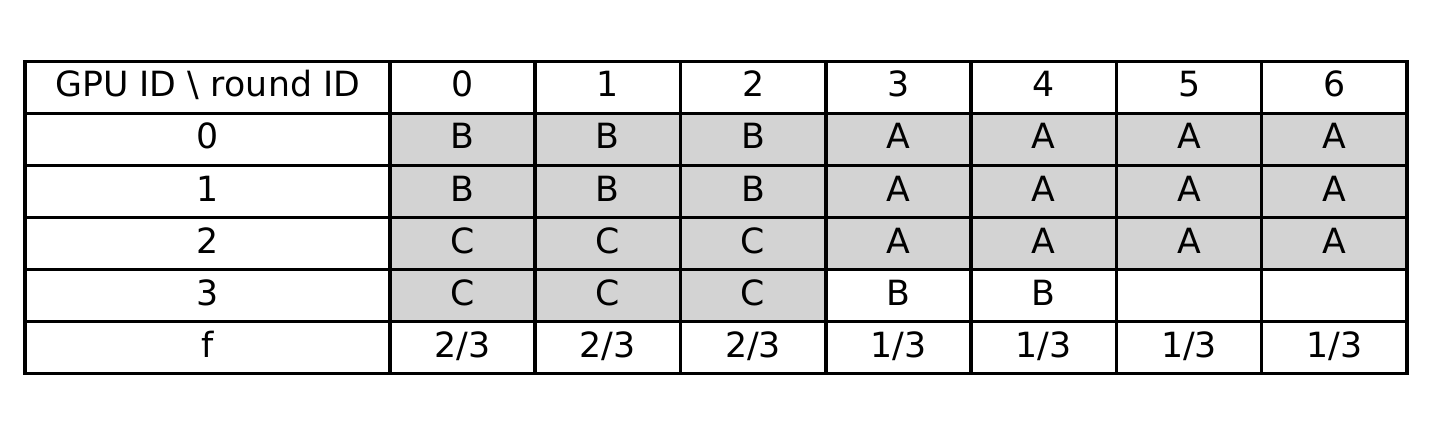}%
      \vspace{-5mm}
    }
    \vspace{3mm}
    \caption{Visualizations of schedules produced by using different filters in Table~\ref{tab:example_filter_1}. The cluster and job setting are the same as those in Figure~\ref{fig:themis-1/3-example}.}
\end{figure}
\section{Volatile Fisher Market (VFM)}\label{apdx:vfm_formulation}
\subsection{Market formulation}
VFM is a dynamic market and continues through rounds indexed by $t=1,\dots,T$. Each round is a fixed time interval, e.g., 120s. Within each round, a central seller (i.e., the scheduler) sells multiple types of resources (e.g., GPUs and/or CPUs) to buyers (i.e., the contending jobs). \footnote{VFM supports multiple-resource allocation, but evaluation in this paper is carried out only for GPU allocation.} All resources are volatile. Resources bought by a job in round $t$ cannot be carried over to future rounds. There is a dynamic price for each resource type in each round, and each job is endowed with an initial budget to spend across rounds. The amount of endowment reflects the priority of jobs. VFM assumes divisible resources \cite{caragiannis2019unreasonable}. 

\noindent \textbf{(a) Buyers, Seller, and Resources.} There exist $N$ buyer jobs competing for $J$ different types of resources. \textbf{(b) Allocation (Purchase).} Let $x_{ijt}$ denote the allocation (purchase) of job $i$ for resource $j$ in round $t$. The resource provision is normalized to one unit. For brevity, let $\bm{x_{it}}$ denote the allocation vector $[\dots,x_{ijt}\dots]$ for job $i$ in round $t$, and let $\bm{X_i}$ denote the $J\times T$ allocation matrix for job $i$ over rounds, with rows and columns corresponding to the resource types and round indices. \textbf{(c) Budget, Price, and Payment.} Job $i$ is endowed with a budget $B_i$ to spend over rounds. The price for resource $j$ in round $t$ is $p_{jt}$; The accrued payment over rounds for job $i$ is $\sum_{j,t}p_{jt}x_{ijt}$. Let $\bm{P}$ denote a $J\times T$ price matrix, with $\bm{P}[j,t]=p_{jt}$. \textbf{(d) Performance (Utility) Function for Dynamic Adaptation.} We use the performance (utility) function $U_{it}(\bm{x_{it}})$ to map received resources, that is, $\bm{x_{it}}$, to the performance gain of job $i$ (e.g., epoch progress). Note that $U_{it}$ can be different over rounds to model time-variant performance under dynamic adaptation. We limit performance functions in the CES family \cite{branzei2014fisher}, which are extensively used in system research.\footnote{Themis \cite{themis20} and Gavel \cite{gavel20} uses linear utility, Dominant Resource Fairness (DRF) \cite{drf11} uses Leontief utility, and REF (Resource Elasticity Fairness) \cite{ref2014} uses Cobb-Douglas utility, which are all CES utility functions.}

\subsection{Solving Equilibrium of VFM}
The market equilibrium captures the optimal allocation for the $N$ jobs in each round, i.e. $\bm{X_1}^*,\allowbreak\dots,\bm{X_N}^*$, and the optimal prices $\bm{P}^*$ for different types of resources in each round. An equilibrium is established if the following two properties are satisfied. \textbf{(a) Maximized Performance Under Budget Constraint (Optimal Spending)}: Each job's performance is maximized under budget constraints. $\bm{X_i}^*=\argmax_{\bm{X_i}} U_i(\bm{X_i}), s.t., \sum_{t}\sum_{j}p_{jt}x_{ijt}\leq B_i,\;\forall i$. \textbf{(b) Work-conserving (Market Clearing)}: There is no leftover resource if the price for the resource type is non-zero. That is, if $p_{jt}>0$, then $\sum_{i}x_{ijt}=1,\;\forall j,t$. 
\begin{theorem}\label{theorem_eg2equilibrium}
For Volatile Fisher Market with linear or Leontief (e.g., DRF\cite{drf11}) utility, the solution of (\ref{eq:egprog}) captures the optimal allocation in the market equilibrium and the Lagrangian dual to capacity constraints (i.e., $\sum_i x_{ijt}\leq 1,\;\forall j,t$) captures the equilibrium price.
\end{theorem}

\begin{equation}\label{eq:egprog}
\begin{gathered}
  \underset{\bm{X_1},\dots,\bm{X_N}}{{Maximize}}\sum_i B_i\, log\, U_i(X_i)\;\; s.t.,%
   \;U_i(X_i)=\sum_t u_{it}(\bm{x_{it}}), \forall i,\\
   \begin{cases}
       u_{it}(\bm{x_{it}})=\sum_{j}u_{ijt}x_{ijt},\forall i,t\, \text{(Linear)}\\
       u_{it}(\bm{x_{it}})=min_{j}\frac{x_{ijt}}{a_{ijt}},\forall i,t\, \text{(Leontief)}
      \end{cases}
      ,\;\;\\
      \sum_{i} x_{ijt}\leq 1,\;\forall j,t,
      \;\;x_{ijt}\geq 0,\;\forall i,j,t
\end{gathered}
\end{equation}

For VFM with linear and Leontief performance function at each instant, Theorem~\ref{theorem_eg2equilibrium} states that the solution of an Eisenberg-Gale \cite{branzei2014fisher} styled program defined in \eqref{eq:egprog} captures the market equilibrium (cf., proof in Appendix \ref{apdx:theorem_eg2equilibrium}). \textbf{Note that even if the instantaneous utility is Leontief, the summed utility over time is not Leontief in general.}

\section{Proof of Theorem~\ref{theorem_eg2equilibrium}}\label{apdx:theorem_eg2equilibrium}
\subsection{Linear Utility}

Volatile Fisher Market (VFM) with linear utilities reduces to a special case of static Fisher market, if we consider volatile resource $j$ at each different time $t$ a unique type of resource. Upon substituting the tuple of resource and time index ($j$, $t$) with a new resource index $k$ (i.e., $(j,t)\rightarrow k$), the Eisenberg-Gale program defined in \eqref{eq:egprog} is equivalent to the program \eqref{eq:egprog_classic}. 
\begin{equation}\label{eq:egprog_classic}
\begin{gathered}
    \underset{\bm{x}}{Maximize}\sum_i B_i\, log\, u_i(\bm{x_i})\;\; s.t.\\
   \qquad u_i(\bm{x_i})=\sum_{k}u_{ik}x_{ik} \\
   \qquad \sum_{i} x_{ik}\leq 1,\;\forall k\\
    x_{ik}\geq 0,\;\forall i,k
\end{gathered}
\end{equation}
Existing work \cite{branzei2014fisher} has proven that program (\ref{eq:egprog_classic}) captures the market equilibrium of a static Fisher market, and thus, it also captures the market equilibrium of the equivalent VFM.

\subsection{Leontief Utility} 

However, VFM with Leontief utilities has no direct link to the classic static Fisher market. We prove the Eisenberg-Gale (EG) program defined in \eqref{eq:egprog} captures market equilibrium by characterizing the Karush–Kuhn–Tucker (KKT) \cite{gordon2012karush} conditions. We rewrite~\eqref{eq:egprog} in a standard convex optimization form and number the constraints as follows:

\begin{subequations}\label{eq:egprog_leontief}
\begin{align}
    \underset{\bm{X_1},\dots,\bm{X_N}}{Minimize}-\sum_i B_i\, log\, U_i(X_i)\;\; s.t.\\
   \qquad U_i(X_i)\leq\sum_t u_{it}(\bm{x_{it}}),\;\forall i \label{eq:eg_const1}\\ 
   u_{it}(\bm{x_{it}})\leq x_{ijt}/a_{ijt},\;\forall i,j,t \label{eq:eg_const2}\\
   \qquad \sum_{i} x_{ijt}\leq 1,\;\forall j,t \label{eq:eg_const3}\\
    x_{ijt}\geq 0,\;\forall i,j,t \label{eq:eg_const4}
\end{align}
\end{subequations}

Let $\beta_{i}$, $\lambda_{it}$, $p_{jt}$, $\eta_{ijt}$ denote the Lagrangian multipliers corresponding to constraints \eqref{eq:eg_const1}, \eqref{eq:eg_const2}, \eqref{eq:eg_const3},\eqref{eq:eg_const4}, respectively. The Lagrangian dual function is

\begin{align*}
L(x,\bm{\beta},\bm{\lambda},\bm{p},\bm{\eta})=-\sum_{i}B_i\,log\, U_i(X_i)+\\\sum_{i,j,t} (u_{it}(\bm{x_{it}})-\frac{x_{ijt}}{a_{ijt}})\lambda_{ijt}\\
+\sum_i(U_i(\bm{X_i})-\sum_t u_{it}(\bm{x_{it}}))\beta_i+\\\sum_{j,t}(\sum_ix_{ijt}-1)p_{jt}-\sum_{i,j,t}x_{ijt}\eta_{ijt}
\end{align*}

First, KKT requires a first-order condition of the Lagrangian function; the gradients to all primal variables and Lagrangian multipliers should be zero. This implies that
\begin{align*}
    &\frac{\partial L}{\partial U_i(\bm{X_i})}=0\implies -\frac{B_i}{U_i(\bm{X_i})}+\beta_i=0\implies \beta_i=\frac{B_i}{U_i(\bm{X_i})}\\
    &\frac{\partial L}{\partial u_{it}(\bm{x_{it}})}=0\implies \sum_j\lambda_{ijt}-\beta_i=0\implies \beta_i=\sum_j\lambda_{ijt}\\
    &\frac{\partial L}{\partial x_{ijt}}=0\implies -\frac{\lambda_{ijt}}{a_{ijt}}+p_{jt}-\eta_{ijt}=0
\end{align*}

The combination of the first two equations implies that
\begin{align*}
  \sum_j\lambda_{ijt}=\frac{B_i}{U_i(\bm{X_i})},\; \forall i,t
\end{align*}

Lagrangian multipliers are nonnegative, and thus, $\eta_i\geq 0$ implies
\begin{align*}
     p_{jt}a_{ijt}\geq \lambda_{ijt}
\end{align*}

Combining the last equation and the last inequality implies that
\begin{align*}
     \sum_j p_{jt}a_{ijt}\geq \sum_j\lambda_{ijt}=\frac{B_i}{U_i(\bm{X_i})},\; \forall i,t\\
     \implies U_i(\bm{X_i})\geq \frac{B_i}{\sum_j p_{jt}a_{ijt}},\; \forall i,t\\
      \implies U_i(\bm{X_i})= \frac{B_i}{min_{t'}\{\sum_j p_{jt'}a_{ijt'}\}},\; \forall i
\end{align*}

This states that for any job $i$, its overall utility is achieved by purchasing resources only in certain time periods that guarantee \textbf{MBB (Maximal Bang-Per-Buck)} \cite{branzei2014fisher} for the job, where $\sum_j p_{jt}a_{ijt}$ represents the unit cost to obtain one unit of utility. MBB guarantees that any job's utility accrued over time is maximized given a fixed budget $B_i$. 

We have proved optimal spending (i.e., maximized utility under budget limit) for each job at the solution of program \eqref{apdx:theorem_eg2equilibrium}. The last step is to prove market clearing.
KKT conditions also require complementary slackness:
\begin{align*}
     p_{jt}(\sum_ix_{ijt}-1)=0,\; \forall j,t\implies\\ \text{if }p_{jt}>0 \text{ then }\sum_ix_{ijt}=1,\; \forall j,t
\end{align*}

This implies that if a resource $j$ is traded in time period $t$, then it must be exhaustively allocated to the jobs and the amount of leftover resources is zero. Therefore, we prove \textbf{MC (Market Clearing)}. An extra fact implied by the solution of the EG program \eqref{apdx:theorem_eg2equilibrium} is that since the objective is to maximize social welfare (budget-weighted geometric mean of jobs' utilities), and thus, there should be no money left by a job in the solution of \eqref{apdx:theorem_eg2equilibrium}. This proves \textbf{BC (Budget Clearing)}. That is, the budget for all jobs will be completely burnt over periods. 

In summary, proving Maximal Bang-Per-Buck, Market Clearing and Budget Clearing establishes the VFM market equilibrium produced by solving program (\ref{eq:egprog_leontief}).
\section{Proof of Theorem~\ref{theorem_ftf}}\label{apdx:theorem_ftf}
\textit{Proof} of (a): $\underset{i}{\prod} \rho_i$= $\underset{i}{\prod} U_i(\frac{C}{N})\cdot \underset{i}{\prod}U_i(\bm{\frac{\bm{X_i}}{N}})^{\frac{-B_i}{B}}$=$\prod_iU_i(\frac{C}{N})\allowbreak \mathsf{NSW_{OT}}^{-1}$. Since $\prod_i\allowbreak U_i(C/N)$ is a constant independent of $X_i$, VFM equilibrium that maximizes $\mathsf{NSW_{OT}}$ equivalently minimizes the product of FTF (Finish Time Fairness) metrics over all jobs, i.e., $\prod_i \rho_i$.

\textit{Proof} of (b). At VFM equilibrium, any job $i$ has maximized utility under budget. When all job have an equal budget, job $i$ will not prefer any other jobs $j$'s allocation, since job $i$ can afford to buy any other job's allocation under same budget. Formally, we get that $U_i(\bm{X_{i}})\geq\allowbreak U_i(\bm{X_{j}})\allowbreak,\, \forall i,j$. Since the market clears in VFM equilibrium, it is not possible that all jobs have a strictly smaller resource share than $C/N$, and there must exist a job $k$ such that its resource share is greater than or equal to $C/N$, then we know $U_i(\bm{X_{j}})\allowbreak\geq U_i(\bm{X_{k}})\allowbreak\geq U_i(\frac{C}{N})$, and thus, Finish Time Fairness if proved.
\section{Stochastic Dynamic Program for Efficiency and Fairness in Expectation}\label{apdx:egprogram_stochastic}

\textbf{(a) State.}  Each job has a private, finite set of states. For dynamic scaling of the batch size, a state is a tuple $(\mathsf{Batch Size}, \mathsf{Epoch})$, which denotes the current batch size and the current epoch (index).  Let $s_{it}$ ($\bm{s_{t}}$) denote the state of job $i$ in round $t$. \textbf{(b) Policy.} Let $\bm{x_{it}}$ ($\bm{x_{t}}$) denote the resource allocated to job $i$ in round $t$. An allocation policy $\pi(\bm{s_{t}},\bm{x_t})$ indicates the probability of making allocation $\bm{x_t}$ to the jobs, conditional on job states $\bm{s_{t}}$, in round $t$. We further limit $\pi$ to be a deterministic policy in this study. \textbf{(c) Transition Probability.} We model the state transition with a probability matrix $P_i(s_{it+1}|s_{it}, \bm{x_{it}})$, which indicates the probability of job $i$ transitioning to state $s_{it+1}$ from state $s_{it+1}$, under resource allocation $x_{it}$. Let $P(\bm{s_{t+1}}|\bm{s_{t}}, \bm{x_{t}})$ denote the transition probabilities for all the jobs. \textbf{(d) Performance (Utility) Function for Dynamic Adaptation.} Let $U_i(s_{it},\cdot, s_{it+1})$  denote the performance gain (e.g., epoch progress) of job $i$ when transitioning from state $s_{t}$ to the next state $s_{t+1}$. Let $U_i(\bm{s_t}, \cdot, \bm{s_{t+1}})$ denote the performance function for all jobs.  

\textbf{Maximized Nash social welfare in expectation.} We construct a linear program in \eqref{eq:seq_eg} to search for an optimal policy that maximizes Nash social welfare in expectation, i.e., $\mathsf{NSW}_{OTE}$. Maximized $\mathsf{NSW}_{OTE}$ co-optimizes efficiency and fairness in expectation sense. The first constraint in \eqref{eq:seq_eg} defines expected cumulative utility under the policy; The second constrains the summed allocation at each period $t$ not exceeding resource provision, and allocation should be non-negative. The third constrains valid probability transition between states. Other constraints are omitted.
\begin{align}\label{eq:seq_eg}
    \pi^*=\underset{\pi}{argmax} \sum_i B_i\, log\,\E_\pi [U_i],\; s.t.\\
    \E_\pi[U_i]=\sum_{t=1}^{T-1}\sum_{\bm{s_t}\in S} \sum_{\bm{x_t}\in X} \sum_{\bm{s_{t+1}} \in S} [\pi(\bm{s_t},\bm{x_t}) \cdot P(\bm{s_{t+1}}|\bm{s_t},\bm{x_t})\nonumber\\
    \cdot U_i(\bm{s_t}, \bm{x_t}, \bm{s_{t+1}})], 
    \;\pi(\bm{s_{t}},\bm{x_{t}})\geq 0 \text{ and } \pi(\bm{s_t},\bm{x_t})=0\nonumber\\
    \text{ if } ||x_t||_{\mathscr{l}_1}>1 \text{ or } x_t<0,\;\forall  \bm{s_t}\in S,\forall \bm{x_t}\in X, t=1,\dots, T\nonumber \\[3pt]
   \sum_{\bm{x_1}}\pi(\bm{s_1},\bm{x_1})=b(\bm{s_1}),\nonumber\\ 
   \sum_{\bm{x_{t}}}\pi(\bm{s_{t}},\bm{x_{t}})=\sum_{\bm{s_{t-1}}}\sum_{\bm{x_{t-1}}} P(\bm{s_{t}}|\bm{s_{t-1}},\bm{x_{t-1}})\pi(\bm{s_{t-1}},\bm{x_{t-1}})  ,\;\nonumber\\t=2,\dots,T\nonumber
\end{align}

\section{\project Design Details}\label{apdx:shockwave_design}
\project plans the schedule for a configurable number ($T$) of future rounds (default $T$: 30 two-minute rounds) and recomputes the schedule when the planned rounds elapse or when jobs arrive or complete. \project's solved schedule is a $N\times T$ binary matrix $\mathsf{X}[j,t]$. $N$ is the total number of active jobs available for scheduling. $\mathsf{X}[j,t]=1$ ($\mathsf{X}[j,t]=0$) represents scheduling (descheduling) job $J_j$ in round $t$ ($t=1,\dots,T$). We next describe the logic in one shot of schedule solving.

\noindent \textbf{Decomposing job schedules to regime schedules.} If dynamic adaptation is predicted to occur within the future planning window, the scheduler must incorporate dynamic changes of jobs' throughputs when solving the schedule.

\textit{Example - A job's dynamic adaptation process has two regimes. The job is currently in the first regime at epoch 5 and the scheduler predicts that the second regime will start from epoch 15. Suppose the planning window is 30-minute long and the epoch duration for the first and second regime are 2 minutes and 1 minute, respectively. Then dynamic adaptation can start as early as the 20th minute in the window, and a $2\times$ change in throughput should be concerned.}

To support dynamic changes in job throughputs, we decompose a job's schedule into its regimes' schedules, such that each regime is a micro-job with static throughput.  In our above example, epochs 5 to 14 will be one micro-job while epoch 15 onward will be the second micro-job. We build a $K\times T$-dimensional binary matrix $\mathsf{Y}_j[k,t]$ to represent the schedule of job $J_j$'s $K$ regimes that can fit in the planning window. $\mathsf{Y}_j[k,t]=1$ ($\mathsf{Y}_j[k,t]=0$) indicates scheduling (descheduling) the $k$-th regime of job $J_j$ to the $t$-th round in the window. Note that partial order constraints are needed to preserve the sequential order between regimes.

\subsection{Implementing Nash Social Welfare over Time}

\noindent We compute the utility of job $J_j$ under schedule $\mathsf{Y}_j[\cdot,\cdot]$ as: 
\begin{align}
\mathsf{UTIL}_j(\mathsf{Y}_j[\cdot,\cdot])=\cfrac{F_j}{E_j}\;+
\sum_{t=1}^{T} \sum_{k=1}^{K} \cfrac{\mathsf{Y}_j[k,t] \cdot D\cdot \mathsf{TH}(j,k)}{Q_j\cdot E_j}
\end{align}
Job $J_j$'s utility equals its current epoch progress percentage (num. finished epochs $F_j$ divided by total num. epochs $E_j$), plus the resulting epoch progress percentage under allocation; the latter sums up the progress percentages for the job across regimes and rounds in the window.\footnote{The epoch progress at a single round $t$ for the $k$-th regime equals the duration of each round ($D_j$) times if the regime is scheduled to the round ($\mathsf{Y}[k,t]$), then divided by epoch duration  $\mathsf{Q}(j)/\allowbreak \mathsf{THPT}(j,k)$}. At the cluster level, the (logarithm of) Nash social welfare over time ($\mathsf{NSW_{OT}}$) integrates the utilities of individual jobs. $\mathsf{Y}[\cdot, \cdot,\cdot]$ is a three-dimensional array that includes the schedule variable for all active jobs' regimes, at all rounds, in the planning window. As stated in \cref{subsec:equilibrium_properties}, maximizing $\mathsf{NSW_{OT}}$ yields an equilibrium and establishes efficiency and fairness guarantees. 
\begin{equation}\
\mathsf{WELFARE}(\mathsf{Y}[\cdot, \cdot,\cdot])= \sum_{j=1}^{N} log\,\mathsf{UTIL}_j(\mathsf{Y}_j[\cdot,\cdot])
\end{equation}
\subsection{Implementing Estimators for Long-Term Effects} 
As previously stated, maximizing social welfare for an (infinitely) long time horizon is difficult due to prohibitive computational overhead and limited predictability. Another reason is that jobs arrive and complete online, and frequent re-planning is unavoidable. In practice, \project only plans the schedule for a finite length window (e.g, 30-60 minutes), and we design estimators that can capture the long-term fairness and long-term efficiency that arise from short-term planning. 

\noindent \textbf{An estimator for long-term fairness.} 
\begin{align}\label{eqn:ftf-heuristic-apdx}
\widehat{\rho}(j)=\frac{L_j+W_j+\mathsf{\widehat{R}(j)}\cdot\mathsf{N_{avg}}(j)}{\mathsf{\widehat{P}(j)}\cdot \mathsf{N_{avg}}(j)}
\end{align}

We estimate the finish time fairness (FTF) $\widehat{\rho}(j)$ of job $J_j$ as its predicted job completion time (the sum of attained service time $L_j$, waiting time $W_j$, and the interpolated remaining run time $\mathsf{\widehat{R}(j)}N_{avg}(j)$), divided by its predicted job run time ($\mathsf{\widehat{P}(j)}N_{avg}(j)$). 

$\mathsf{\widehat{P}(j)}$ ($\mathsf{\widehat{R}(j)}$=$\mathsf{\widehat{P}(j)}-L_j$) is the total (remaining) run time under isolated resources predicted using the Bayesian posterior. Similarly to prior work \cite{themis20}, we linearly scale the isolated run time with a contention factor $N_{avg}(j)$ to compute the run time under contention. In this paper, we define the {\em contention factor} as, within a fixed time range, the ratio between the number of jobs requesting GPUs and the overall number of GPUs provisioned in the cluster, and a job's contention factor $N_{avg}(j)$ only accounts for the time range it is either queued or running.

\project plugs in the $k$-th power of FTF $\rho$s of jobs into social welfare function (see Equation \ref{eqn:schedule-solver-apdx}) as weights. The weights in the social welfare function act as the budgets assigned to jobs in the volatile Fisher market. If a job is predicted to be unfairly scheduled (large FTF $\rho$) in the long term, VFM correspondingly assigns a higher budget for it and proactively prioritizes the job in the planning window.

\noindent\textbf{An estimator for long-term efficiency.}
The efficiency estimator estimates the final makespan to complete all current jobs' training epochs and penalizes schedules (in the planning window) that potentially increase the makespan estimate. However, the final makespan is unknown at the current instant and, in practice, \project penalizes increasing the lower bound of it. \project uses the lower bound given in~\cite{coffman1978application}. Let $\mathsf{R}(\mathsf{Y}_j[\cdot,\cdot])$ denote the remaining run time of job $J_j$ from the planning window. The lower bound of makespan (for the remaining epochs) is estimated as the maximum between the sum of the remaining run time divided by the number of GPUs in the cluster (i.e., $M$), and the longest remaining run time among jobs. Intuitively this takes the maximum between the longest job remaining and the makespan if all remaining jobs were evenly spread out across the cluster.

\begin{align}\label{eqn:makespan-heuristic}
\mathsf{H(\mathsf{Y}[\cdot, \cdot,\cdot])}=max\{\frac{\sum_j \mathsf{R}(\mathsf{Y}_j[\cdot,\cdot])}{M},\, max_j\mathsf{R}(\mathsf{Y}_j[\cdot,\cdot])\}
\end{align}

Finally, we plug in the long-term efficiency estimator to social welfare maximization as a regularizer (See Equation~\ref{eqn:schedule-solver-apdx}). $\lambda$ is a tunable coefficient that controls the degree of regularization. \project yields similar makespan and fairness for different workloads when $\lambda$ is between 1e-1 and 1e1.

\subsection{An End-to-End Schedule Optimizer}
Finally, Equation~\ref{eqn:schedule-solver-apdx} shows the optimization problem solved by \project{} in a given round. The output of the solver is schedule for each regime and the job schedule for the round can be simply translated from the regime schedule.
\begin{align}
\begin{split}\label{eqn:schedule-solver-apdx}
\underset{Y_1, \dots, Y_N}{\mathsf{Maximize}}\; \frac{1}{NM}\sum_{j=1}^{N} {\rho(j)}^{k}log[\,\mathsf{UTIL}_j(\mathsf{Y_j[\cdot,\cdot]})]\\
- \frac{\lambda}{Z_0} \mathsf{H(\mathsf{Y}_1[\cdot,\cdot],\dots, \mathsf{Y}_N[\cdot,\cdot])}
\end{split}
\end{align}
$Z_0$ is a normalization coefficient, which is the sum of the interpolated run time across all jobs. More details about the constraints can be found in Appendix~\ref{apdx:solver_constraints}.

\textbf{Handling dynamic job arrival.} Similar to existing schedulers, such as Themis \cite{themis20}, Tiresias \cite{gu2019tiresias}, Pollux \cite{qiao2020pollux} and Gavel \cite{gavel20}, \project periodically adds newly arriving jobs to the schedule solver (Equation~\ref{eqn:schedule-solver-apdx}). The fairness objective in \project  (Equation~\ref{eqn:ftf-heuristic-apdx}) automatically handles selecting between newly-arrived short jobs or jobs that have been waiting in the queue for a long time, according to their pressure on breaking finish time fairness.

\section{Constraints Of Program \ref{eqn:schedule-solver-apdx}}\label{apdx:solver_constraints}
Program \ref{eqn:schedule-solver-apdx} requires the following constraints (details omitted). \textit{(1) Preserving the order of regimes.} Any regime is prohibited to run before precedent regimes are complete. \textit{(2) Work-conserving (Market Clearing).} Idle resources are not allowed when there are ready jobs. \textit{(3) Capacity Limits.} GPUs assigned to jobs should not exceed the overall provision.
\section{Varying Contention Factor}\label{subsec:vary_contention}

\begin{figure}[!tbp]
 \centering
\includegraphics[width=0.5\textwidth, trim=0.1cm 0.1cm 0.1cm 0.1cm,clip]{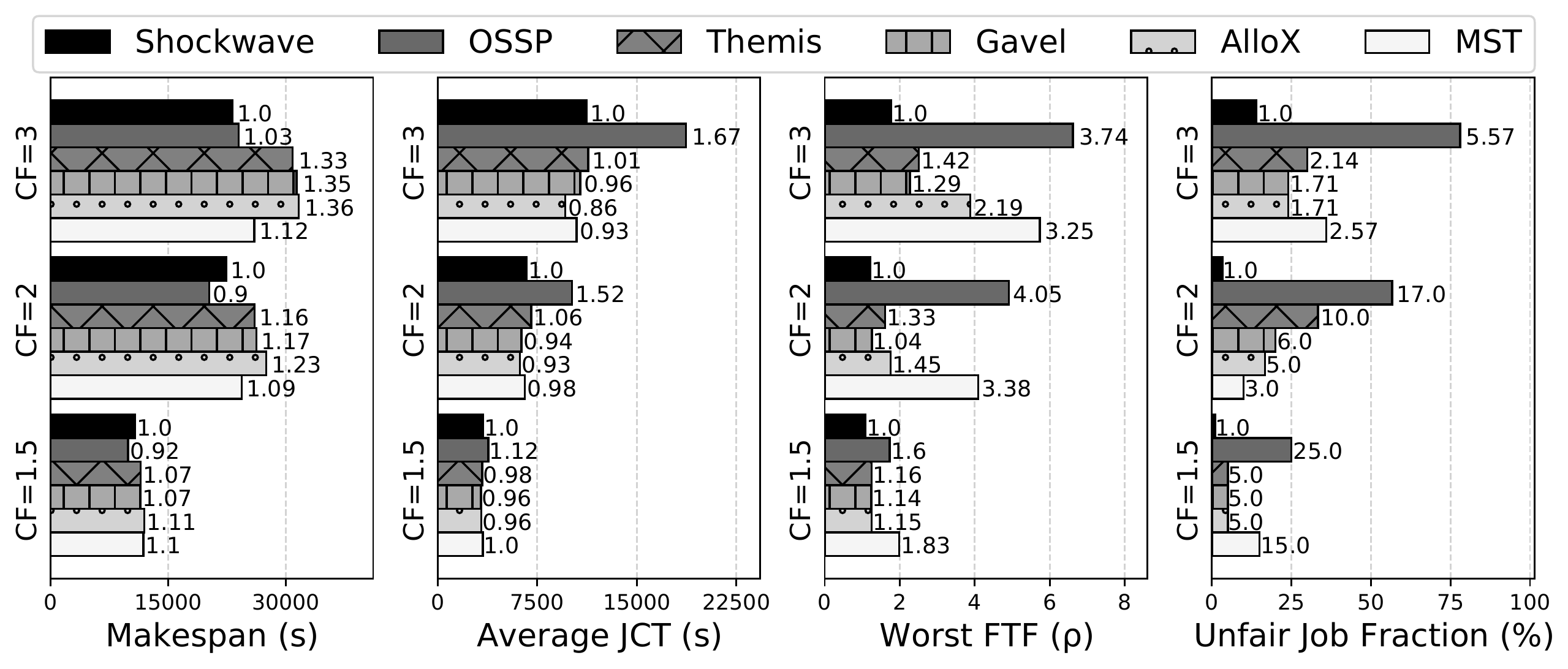} 
\caption{[Physical] Evaluating the scheduling efficiency and fairness of \project under different contention factors (CF) in a 14-GPU physical cluster.}
\label{fig:physical-cfs}
\end{figure}

We define the contention factor as, within a fixed time range, the ratio between the number of jobs requesting GPUs and the overall number of GPUs provisioned in the cluster. A larger contention factor indicates more jobs competing GPU resources at an instant. So far, we have assumed a default contention factor (three). We next vary the contention factor and compare policies on a smaller 14-GPU physical cluster.

\project's win in efficiency decreases as there is more resource slack and less contention in the cluster. \project's improvement in makespan over Gavel, AlloX, and Themis decreases (from 35\% for contention factor 3) to 19\% (8\%) when the contention factor is lowered to 2 (1.5) (cf. Figure~\ref{fig:physical-cfs}). A similar trend for cluster utilization is found. \project's improvement in cluster utilization drops to 19\% (5\%). Although the finish time fairness of all policies improves as the contention factor decreases, \project still performs better than the baselines.  \project keeps the fraction of unfairly scheduled jobs (i.e., the fraction of jobs with FTF $\rho$>1) low when varying the contention factor. The average fraction of unfairly scheduled jobs for \project is 8.67\% when varying the contention factors, outperforming the baselines by 2.85$\times$ (see Figure~\ref{fig:physical-cfs}).  When the contention factor is lowered to 2, \project maintains a worst-case FTF $\rho$ of 1.2, outperforming Themis, Gavel, and AlloX by 1.27$\times$. When the contention factor is further lowered to 1.5, \project and all the baselines worst-case FTF approach 1 and the difference is insignificant.

\section{Varying the Cluster Trace}\label{subsec:eval_pollux_arrival}
\begin{figure}[!tbp]
\centering
\includegraphics[width=0.475\textwidth, trim=0.1cm 0.1cm 0.1cm 0.1cm,clip]{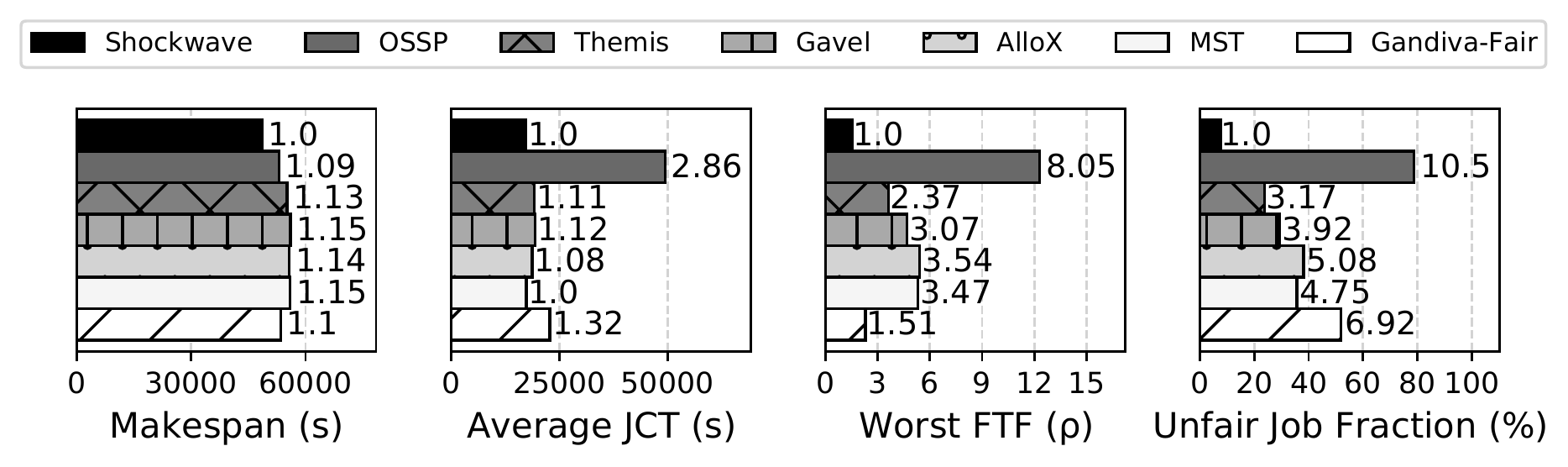} 
\caption{[Simulation] Evaluating \project's scheduling efficiency and fairness for Pollux trace on a 32-GPU cluster.}
\vspace{-0.1in}
\label{fig:pollux_arrival}
\end{figure}
In previous subsections, we presented the results of \project using synthetic traces generated by the Gavel\cite{gavel20} workload generator. In this subsection, we extend the evaluation using real DNN training traces provided by the Pollux \cite{qiao2020pollux} system. The Pollux trace provides the duration and arrival timestamps for training jobs and is extracted from a previous workload analysis~\cite{philly19}.  Figure~\ref{fig:pollux_arrival} shows the comparison between \project and the baseline algorithms and we can see a similar trend as in previous sections. However, the win in makespan over Themis, Gavel, and AlloX drops from 30-35\% to 20\% on the Pollux trace. In previous synthetic traces, the duration of jobs has a greater diversity ($2\times$) than in the Pollux trace, and thus long-running jobs have a larger impact on final makespan and cluster utilization. Therefore, opportunistically prioritizing these long-running jobs leads to greater improvement when there is more diversity among jobs.

\end{document}